\documentclass[a4paper,11pt]{article}
\usepackage{epsfig}
\usepackage{amsfonts}
\usepackage{latexsym}
\usepackage{color}
\usepackage{amsfonts}
\usepackage{graphicx}
\usepackage[colorlinks]{hyperref}
\addtocounter{section}{0}

\newcommand{\ba}{\begin{array}}
\newcommand{\ea}{\end{array}}
\newcommand{\be}{\begin{equation}}
\newcommand{\ee}{\end{equation}}
\newcommand{\nn}{\nonumber}
\newcommand{\bea}{\begin{eqnarray}}
\newcommand{\eea}{\end{eqnarray}}
\newcommand{\beas}{\begin{eqnarray*}}
\newcommand{\eeas}{\end{eqnarray*}}

\def\be{\begin{equation}}
\def\ee{\end{equation}}
\def\ba{\begin{array}}
\def\ea{\end{array}}
\def\bea{\begin{eqnarray}}
\def\eea{\end{eqnarray}}
\def\dd{\partial}

\def\a{\alpha}
\def\b{\beta}

\def\e{\varepsilon}

\begin{document}

\begin{center}
\begin{Large} {\Large{
New solutions to the $s\ell_q(2)$-invariant Yang-Baxter equations
at roots of unity: cyclic 
 representations}}
\end{Large}\vspace{0.2cm}\\ {D. Karakhanyan\footnote{e-mail:{\sl karakhan@mail.yerphi.am}}}, {Sh. Khachatryan\footnote{e-mail:{\sl shah@mail.yerphi.am}}} \\
\begin{small}{Yerevan Physics Institute, Br. Alikhanian 2, Yerevan
36, Armenia}\end{small}
\end{center}

 We find the all solutions to the $sl_q(2)$-invariant multi-parametric Yang-Baxter equations (YBE) 
 at $q=i$
 defined on the cyclic (semi-cyclic, nilpotent)
 representations of the
 algebra. We are deriving the solutions in form of the linear combinations over the
 $sl_q(2)$-invariant objects - projectors. The direct
 construction of the projector operators at roots of unity
 gives us an opportunity to consider all the possible cases, including
  also  degenerated
 one, when the number of the projectors becomes larger, and various type of solutions are arising, and  as well
 as the inhomogeneous case. We are giving a full
 classification of the YBE solutions for the considered representations. A
 specific character of the solutions is the existence of the
 arbitrary functions.

{\small\tableofcontents}

 \section{Introduction}

The study of the representation theory of the quantum algebras at
roots of unity was begun 
at the early nineties of the past century \cite{s1,ps,ak}. At the
same time  it stimulated the works on the investigation of the
(universal) intertwiner $R$-matrices defined on the non-standard
representations (which have no analogies in the case of
non-deformed algebra). Such matrices, as well as  intertwiner
matrices for  affine extensions of the quantum algebras were
constructed and there were observed the models possessing the
quantum algebra symmetry at roots of unity
\cite{ak,asc,AuP,RVJ,BS,RGS,RGSs,grs,DA}. Especially
 the simplest case of the quantum algebra  $sl_q(2)$ at roots of unity ($q^N=1$)
 was
thoroughly investigated. The connection of the Potts model's
$R$-matrix  with the intertwiner $R$-matrices defined on the
cyclic representations of the algebra $\hat{sl}_q(2)$ ($N\geq 3$)
was observed \cite{AuP,BS}. The $R$-matrices defined on the
semi-cyclic (nilpotent) representations of $sl_q(2)$ were explored
in \cite{RGSs}. As it is known the intertwiner matrices satisfy
the Yang-Baxter equations (YBE) \cite{ybe,ks}. Solutions to the
YBE with the non-standard representations are investigated in the
series of the papers \cite{RVJ,RGS,RGSs,DA} and some explicit
solutions are obtained. However we think that the solutions to the
Yang-Baxter equations with the
 cyclic, semi-cyclic and nilpotent, as well as
  indecomposable representations of the quantum
 algebra $sl_q(2)$ at roots of unity  need a thorough investigation.
 There is known the decomposition of the intertwiner matrices over the
 symmetry-invariant objects - projectors \cite{krs,j}. For constructing the
 projectors explicitly at first one has to determine the fusion rules
  at roots of unity \cite{ak}.  Using the detailed rules, formulated in \cite{kkh} for
   the highest/lowest weight indecomposable representations,
   in our previous paper \cite{shkh} by means of the
 direct
construction of the projection operators, we see that the
 consideration  
 of the highest/lowest weight indecomposable representations even
 for the simplest case $q^4=1$ gives a large amount of
 various new solutions. Considering
 the whole set of
 the projection operators 
  we ensure  
  the foundation of the all  possible solutions for the given representations.

  Here we investigate the YBE  with the $sl_q(2)$-invariant
 $R$-matrices, defined on the cyclic (semi-cyclic, nilpotent)
 irreps, again at $q^4=1$, which means that we work with $4\times
 4$ matrices. And now also we find rich variety of  solutions.
 As at roots of unity the center of the algebra is enlarged and
 the cyclic representations are parameterized by means of the
 continuous parameters (in addition to the  eigenvalues
  of the quadratic Casimir operator), such parameters are involved
  in the YBE as new parameters, and in general here we deal with
  the multi-parametric YBE.  We would like to
  emphasize, that  the case of $q^4=1$  was investigated
  in \cite{RGS}, where there were obtained particular solutions,
 with the matrix elements connected with the Clebsh-Gordan coefficients.
  The mentioned work contains
  first hint about a remarkable property of the general
 solutions defined on the cyclic irreps at $q^4=1$, that is the existence
 of the arbitrary
 functions. Therein the author noted that the obtained
 solutions do not exhaust the all list of  possible solutions at
 $q^4=1$. In  \cite{DA} the authors have constructed $R$-matrices
 defined on the $N$-dimensional irreps of $sl_q(2)$ algebra at roots of unity
 $q^{2N}=1$ ($N$-state colored braid matrices), taking the appropriate limit of $q$ from the
   YBE solutions defined on the infinite dimensional representations at general
   $q$. The matrices are represented via the Clebsh-Gordan coefficients and are trigonometric functions on the
   arguments.
    For the case $N=2$ this solution corresponds the mentioned
    solution brought in \cite{RGS}, if to set the
    arbitrary functions as trigonometric ones. However, as at
    roots of unity the representation spectra and the fusion rules
    are changed radically, the use of the limits of the formulas
    obtained at general $q$ can provide us only with the part of the
    solutions; the whole set of solutions can
     be obtained if to construct the states
    and projectors directly for the exceptional values of
    $q$ \cite{shkh}, as there can be degenerated situations, when
    the number of the projection operators becomes larger,
    compared to the cases at general $q$.
 We think that the presented  technique allows us to pretend the full
spectra of the YBE solutions defined on  two-dimensional cyclic
irreps. The investigation of the solutions by direct constructions
with the cyclic
  (as well as the indecomposable) representations at higher
 roots of the unity we intend to perform in the further works.

Among the obtained solutions there are entirely new solutions
(presented in the subsection 4.3) and also there are such ones,
which coincide with the already obtained solutions
\cite{RGS,grs,KS}, such as the solution (\ref{ru0})
\cite{RGS,grs,DA} or the solutions (\ref{hg}, \ref{gh}),
 which are the
particular trigonometric  limits of the solutions presented in
\cite{KS,Felder,BSf}
 (see also the citations brought therein), and (\ref{ruw})
 \cite{KS}.
 Thus we unveil the  underlying $sl_q(2)$-symmetry of the mentioned solutions
(\ref{hg}, \ref{gh}, \ref{ruw}).

 All the obtained solutions have the so-called "free-fermionic" property
 \cite{WuChan,Felder,BSf},
 which is the peculiarity of the case $q^4=1$. The corresponding
 quantum one-dimensional spin-chain models are the generalizations of the
 $XY$ model in a transverse magnetic field. This is an expected result, as
 it is known that the free-fermionic $XX$ model corresponds to the case $q=\pm i$
 of the $sl_q(2)$-invariant $XXZ$ model, and also there a correspondence
 is established between the checkerboard $2d$ Ising model (the $N=2$ analog of the chiral Potts model)
  and the free-fermionic $XY$ ($XZ$) models \cite{AuP,rbaxter}. In \cite{DA} it is
  stated the correspondence of the obtained
  $R$-matrix at $N=2$ with the trigonometric limit of the tree-parametric (or
  colored) free-fermionic YBE solutions \cite{Felder,BSf}.  The
    connection of this matrix with the quantum algebras $gl_q(1|1)$ and
    $sl_q(2)$ are shown in \cite{KRS} and \cite{Murakami}.

The paper is organized as follows. In the Section 2 the definition
of the quantum algebra $sl_q(2)$ and it's representations are
brought. The functional representation of the algebra by means of
theta functions is constructed for the cyclic (semi-cyclic,
nilpotent) irreps. 
The polynomial representation for the highest/lowest weight irreps
can be found e.g. in \cite{kkh}. In the Section 3 the YB equations
for two-dimensional cyclic irreps at $q=i$ (all the results can be
extended for the  equivalent case of $q=-i$) are formulated, and
the general aspects of the investigation by means of the
projection operators are explained. In the Section 4 the solutions
to the YBE are presented. In the Section 5 the corresponding
spin-chain quantum models in general terms are sketched and the
summary of the work is given.

\section{Algebra and notations}
\addtocounter{section}{0}\setcounter{equation}{0}

The quantum algebra $s\ell_q(2)$ is defined by the generators
$e,\;
f,\; k^{\pm 1}$ \cite{ak,grs} 
 \be\label{alg}
kek^{-1}=q^2e,\qquad kfk^{-1}=q^{-2}f,\qquad
[e,f]=\lambda^{-1}(k-k^ {-1}),\qquad\lambda=q-q^{-1}.
 \ee
 The quadratic Casimir operator is written as
  \be\label{cas}
c=ef+\frac{q^{-1}k+qk^{-1}}{\lambda^2}. \ee
 At the exceptional values
of $q$ ($q^{N}=1$) the center of algebra is enlarged and three new
Casimir operators appear: $k^{\mathcal{N}}$, $e^{\mathcal{N}}$ and
$f^{\mathcal{N}}$, here ${\mathcal{N}}=N$ if $N$ is odd and
$\mathcal{N}=N/2$ if $N$ is even \cite{s1,ak}. One can check this
by direct calculations of the corresponding commutators.
So the representations are characterized by means of the values of
the mentioned operators 
 \be\label{ce}
e^{\mathcal{N}}=\mathrm{x}{\mathbb{I}},\qquad
f^{\mathcal{N}}=\mathrm{y}{\mathbb{I}},\qquad (k^\pm
1)^{\mathcal{N}}=\mathrm{z}^{\pm 1}{\mathbb{I}} \quad \mbox{and}
\quad c=\mathrm{c}{\mathbb{I}}.
 \ee
 %
The values of the Casimir operators are connected by a relation
(\ref{constr}) \cite{ak}, which will be presented further in this
section. The representations are  grouped into two classes:
$A$-type representations, having highest and lowest weights, which
include usual spin-representations $V_r$ (typical to the algebra
$sl(2)$) with the dimensions $r\leq \mathcal{N}$ and the $2
\mathcal{N}$-dimensional indecomposable representations
$\mathcal{I}_{A}$, arising in the fusions of the spin irreps, and
the $B$-type representations, including $ \mathcal{N}$-dimensional
cyclic (semi-cyclic, nilpotent) irreps $V_{\mathcal{N} }$ and the
corresponding $2 \mathcal{N}$-dimensional indecomposable
representations $\mathcal{I}_B$. For the detailed classification
see \cite{ak}.

 Let us present here the general cyclic irrep
 $\{v_1,\;v_2\;\cdots;v_{\mathcal{N}}\}$,
  $v_{i+\mathcal{N}}\equiv v_i$ at $q^{\mathcal{N}}=\pm 1$ with the action
   of the algebra generators:
\bea k \cdot v_i&=&q^{\varepsilon+2i}v_i,\nn\\
e\cdot v_i&=&\beta_i v_{i+1},\label{rep}\\
 f\cdot v_{i}&=&\gamma_i v_{i-1},\nn\eea
The algebra relations give
 \bea \beta_{i-1}\gamma_{i}-\gamma_{i+1}\beta_{i}=[\varepsilon+2i]_q,\;
\prod_{i=1}^{\mathcal{N}}\beta_i=\mathrm{x},\;
 \prod_{i=1}^{\mathcal{N}}\gamma_i=\mathrm{y},\;q^{\mathcal{N}\varepsilon}=\mathrm{z}.
\eea
The parameters $\beta_i,\;\gamma_i$, connected with the above
equations, can be fixed by normalization conditions. Denoting
$\alpha_i=\gamma_{i+1}\beta_{i}$, we find
$$\alpha_i=\alpha_1-\sum_{p=2}^i[\varepsilon+2p]_q=\alpha_1-[i-1]_q[1+i+\varepsilon]_q.$$
Parameterizing $\alpha_1$ as follows
$\alpha_1=\left[\frac{3+\varepsilon+\xi}{2}\right]_q\left[\frac{\xi-3-\varepsilon}{2}\right]_q$,
we obtain a compact formula
\be \alpha_i=\left[i+\frac{1+\varepsilon+\xi}{2}\right]_q
\left[\frac{\xi-\varepsilon-1}{2}-i\right]_q. \ee
The semi-cyclic or nilpotent irreps correspond to the choice
$\alpha_{\mathcal{N}}=0$, which gives the values
$\xi=\pm\varepsilon\pm 1+2n\mathcal{N}$ (modulo $2{\mathcal{N}}$).
We can verify that the parameter $\xi$ is connected with the
eigenvalue $\mathrm{c}$ of the quadratic Casimir operator $c$.
Acting by the l.h.s and r.h.s. of the relation (\ref{cas}) on the
vector state $v_{i+1}$, we find
$\mathrm{c}=\alpha_i+\frac{q^{\varepsilon+2i+1}-q^{-\varepsilon-2i-1}}{\lambda^2}=
\frac{q^{\xi}+q^{-\xi}}{\lambda^2}$.

 To relate
the values of the Casimir operators \cite{ak,grs} one can start
from the relation (\ref{cas}) in form:
$$
ef=c-\frac{q^{-1}k+qk^{-1}}{\lambda^2},
$$
acting the l.h.s and r.h.s of it on the states of an
$\mathcal{N}$-dimensional cyclic irrep and multiplying the
results, which in fact will form
 the determinants (invariant quantity) of the corresponding $\mathcal{N}\times
 \mathcal{N}$matrices.
So one will obtain in l.h.s. $\prod_{s=1}^{{\mathcal{N}}}\alpha_s=
 \prod_{s=1}^{{\mathcal{N}}}\gamma_{s} \prod_{s=1}^{{\mathcal{N}}}\beta_s=\mathrm{x}\mathrm{y}$. The result in
r.h.s. one can reformulate using the relation
 \be\label{nphi}
\prod_{k=1}^{{\mathcal{N}}}[\a+k]_q,
 =\lambda^{-{\mathcal{N}}}(q^{{\mathcal{N}}\a+\mathcal{N}(\mathcal{N}+1)/2}+
 (-1)^{\mathcal{N}}q^{-{\mathcal{N}}\a-\mathcal{N}(\mathcal{N}+1)/2})\equiv \Phi(\a).
 \ee
 So we arrive at:
 \be
\prod_{s=1}^{{\mathcal{N}}}\alpha_s=\prod_{s=1}^{{\mathcal{N}}}\left(\mathrm{c}
-\frac{q^{\varepsilon+2s-1}+q^{-2s-\varepsilon+1}}{\lambda
^2}\right)\equiv\prod_{s=1}^{{\mathcal{N}}}\left(\frac{q^{\xi}+q^{-\xi}}{\lambda^{2}}-\frac{q^{\varepsilon+2s-1}+q^{-2s-\varepsilon+1}}{\lambda
^2}\right)=
 \ee
$$
=\prod_{s=1}^{{\mathcal{N}}}\left[\frac{\xi}{2}+\frac12(\varepsilon-1)+s\right
]_q \left[\frac{\xi}{2}-\frac12(\varepsilon
-1)-s\right]_q=\lambda^{-2{\mathcal{N}}}\left(q^{{\mathcal{N}}\xi}+q^{-
{\mathcal{N}}\xi}+ (-q)^{\mathcal{N}}(\mathrm{z}+\mathrm{z}^{-1})
\right),
$$
where  the parametrization 
$\mathrm{c}=\frac{q^{\xi}+q^{-\xi}}{\lambda^{2}}$ is used.
  Thus,
\be \label{constr}
xy=\lambda^{-2{\mathcal{N}}}(q^{{\mathcal{N}}\xi}+q^{-
{\mathcal{N}}\xi}+ (\mp
1)^{\mathcal{N}}(\mathrm{z}+\mathrm{z}^{-1})). \ee

Taking into account the relation (\ref{constr}) the cyclic irreps
have three independent characteristics. Besides of the parameters
$\varepsilon,\; \xi$ in the presented representation space
(\ref{rep}) we can introduce the third independent parameter
$\omega$ by fixing the parameters $\beta_i,\;\gamma_i$ in the
following general way:
$\beta_i=\sqrt{\alpha_i}f(\varepsilon,\xi,\omega,i),\;\;\;
\gamma_i=\sqrt{\alpha_{i-1}}/f(\varepsilon,\xi,\omega,i-1)$, with
a function $f(\varepsilon,\xi,\omega,i)$. Particularly we can take
\bea \beta_i=\left[i+\frac{1+\varepsilon+\xi}{2}\right]_q
[\omega+i]_q,\;\;\; \gamma_i=
\left[\frac{\xi-\varepsilon+1}{2}-i\right]_q/[\omega+i-1]_q.\label{par}\eea
Here the parameters $\varepsilon,\;\xi,\;\omega$ are related by
the constraints (\ref{constr}),
$x=\Phi[\frac{1+\varepsilon+\xi}{2}]\Phi[\omega]$ and
$y=\Phi[\frac{\xi-\varepsilon+1}{2}]/\Phi[\omega]$. In respect to
$q^{\xi}$ and $q^{\omega}$ these constraints are the equations of
the $\mathcal{N}$-th degree and have different solutions of number
$\mathcal{N}$. The solutions with different $\xi$ ($\xi_i=\xi_0+i
$, $i=1,..., \mathcal{N}$) are connected with different values of
the quadratic Casimir operator, while the solutions with different
$w$ ($\omega_n=\omega_0+2 n$, $n=1,..., \mathcal{N}$) are entirely
equivalent.

Any cyclic representation with the given Casimir values
$\{\mathrm{x},\mathrm{y},\mathrm{z},\mathrm{c}\}$ can be
characterized by the quantities
$\{\mathrm{x},\mathrm{y},{\varepsilon},{\xi}_i\}$. The semi-cyclic
irreps with the condition $\alpha_{\mathcal{N}}=0$ can be defined
as follows: $\beta_i={\alpha_i}$ and
$\gamma_i=1+(y-1)\delta_{i,1}$, when $x=0$ and there is a highest
weight ($v_{\mathcal{N}}$); or
$\beta_i=1+(x-1)\delta_{i,\mathcal{N}}$ and $\gamma_i={\alpha_i}$,
when $y=0$ and there exists a lowest weight ($v_1$).

The quantum algebra is characterized by  co-product, definition of
which has  some ambiguity, when we check the consistency of the
co-product with the algebra relations. In the case of the general
values of $q$ the generators on the tensor product of two
representations can be chosen in the following general form:
$$
\Delta[k]=k\otimes k,\qquad \Delta[e]=k^a \otimes e+e\otimes
k^b,\qquad \Delta[f]=k^c \otimes f+f\otimes k^d,
$$
which is obviously consistent with the scale part of the symmetry
(\ref{alg}). Then unwanted terms in the  algebra relations cancel
at $d=-a$, $c=-b$ and $a-b=\pm1$. This
 provides one-parameter
families of the co-products $\Delta$ and $\bar\Delta\equiv P\Delta
P$ ($P$ is a permutation map):
 \be\label{a}
\Delta[k^{\pm}]=k^{\pm}\otimes k^{\pm},\qquad \Delta[e]=k^a
\otimes e+e \otimes k^{a+1},\qquad \Delta[f]=k^{-a-1}\otimes
f+f\otimes k^{-a},
 \ee
 \be\label{a+1}
\Delta[k^{\pm}]=k^{\pm}\otimes k^{\pm},\qquad \Delta[e]=k^a
\otimes e+e\otimes k^{a-1},\qquad \Delta[f]=k^{-a+1}\otimes
f+f\otimes k^{-a}.
 \ee
However, when $q$ takes exceptional values
($q^{\mathcal{N}}=\pm1$) only integer (integer and half-integer)
values of $a$
are acceptable. 
 One can check this statement straightforward in
the following way. If we suppose that the operator $k^a$ satisfies
the algebra relation $k^a e=q^{2 a} e k^a$, then we come to $k^a
e^{\mathcal{N}}= q^{2a {\mathcal{N}}} e^{\mathcal{N}} k^a$. As the
operator $e^{\mathcal{N}}$ belongs to the center, it follows that
$q^{2a {\mathcal{N}}}=1$, i.e. the number $a$ ($2a$) must be
integer if $q^{{\mathcal{N}}}=-1$ ($q^{\mathcal{N}}=1$).

 In the further discussion we use the
 formula (\ref{a+1}) with the value $a=1$. Then the operation
 $\bar\Delta$ corresponds to (\ref{a}) with $a=0$. These two
 operations are connected with the intertwiner matrix $R$ defined
 on the space $V\otimes V$:
\be R\Delta=\bar{\Delta}R.\label{com}\ee
%
%
%
 It occurs that the irreps
(representations), on which the intertwiner is defined, must have
correlated parameters: the values of the extended center are
mutually connected  due to the relations (\ref{com}). For general
$\mathcal{N}$ the elements of the center $e^{\mathcal{N}}$,
$f^{\mathcal{N}}$, $k^{\pm\mathcal{N}}$ have the same co-products
as the generators $e,\; f,\; k^{\pm 1}$:
\bea \label{efk}\Delta[e^{\mathcal{N}}]=k^{\mathcal{N}}\otimes
e^{\mathcal{N}}+e^{\mathcal{N}}\otimes
1,\;\;\;\Delta[f^{\mathcal{N}}] =1\otimes f^{\mathcal{N}}+
f^{\mathcal{N}}\otimes
k^{-\mathcal{N}},\;\;\;\Delta[k^{\pm\mathcal{N}}]=k^{\pm\mathcal{N}}\otimes
k^{\pm\mathcal{N}}. \eea
Implying the relation (\ref{com}) for the elements of the center
 $e^{\mathcal{N}},\;
f^{\mathcal{N}}$ on the tensor product of two cyclic
representations with the characteristics $\{x_i,y_i,z_i\}$ and
$\{x_j,y_j,z_j\}$, we arrive at \cite{ak}
\bea z_i x_j+x_i=x_i z_j+x_j,\;\;\; y_j+y_i
z_j^{-1}=y_i+z_i^{-1}y_j.\label{xyz}\eea

\subsection{Functional representation of the algebra}

The algebra (\ref{alg}) can be realized in terms of
finite-difference operators acting on the space of complex valued
functions as follows:
 \be\label{dop1}
e=q^{\gamma/2}e^t[\dd-\a]_qq^{\epsilon\dd},\qquad
f=q^{-\gamma/2-\epsilon \dd }e^{-t}[\b-\dd]_q,\qquad
k=q^{2\dd-\a-\b}.
 \ee
 The parameter $\gamma$
is related to the rescaling of the generators $e$ and $f$, while
the parameter $\epsilon$ is related to an automorphism $e\to
ek^{\epsilon/2}$, $f\to k^{-\epsilon/2}f$. The parameters $\a$ and
$\b$ are also defined up to common shift. For the spin-irreps the
representation space is isomorphic to the space of polynomials of
$e^t$. Then the half-sum $(\a+\b)/2=\ell$ has sense of the spin of
the representation.

The functional realization for cyclic representations, containing
three independent parameters can be obtained from (\ref{dop1}) by
a transformation: $e'=e\sum_{n=0}^{{\mathcal{N}}-1} e_n
q^{\epsilon_n-n+2n\dd}$, $f'=f\sum_{n=0}^{{\mathcal{N}}-1} f_n
q^{\epsilon_n-n+2n\dd}$, $k'=q^{\chi} k$, with some $e_n,\;
f_n,\;\chi$ which can be defined from the algebra relations. From
the another hand we can simply apply the realization (\ref{par})
to the appropriate chosen functional space. The role of monomials
for the cyclic representations can play the following
theta-functions with characteristics:
 \be\label{theta}
\theta_r(t)=\sum_{n=-\infty}^\infty e^{i\pi\tau(n+\frac r{\mathcal
{N}})^2}e^{t(n{\mathcal {N}}+r)},\qquad r=1,\ldots{\mathcal {N}}.
 \ee
The parameter $\tau$ is specified by one requirement:
$$
Im\tau>0,
$$
ensuring the convergence of theta-series. The functions
(\ref{theta}) form basis in the space of entire functions of order
${\mathcal{N}}$ \cite{skl}.

The cyclic property is implied in this realization by the fact
that shifts induced by derivative $\dd_t$ on $r$ units are defined
by modulo ${\mathcal {N}}$ due to the periodicity of
theta-functions (\ref{theta}). In order to find an operator
realization of generators corresponding to (\ref{par}) acting on
basis (\ref{theta}) one should just replace the parameter $i$ in
the expressions of the matrix elements of generators by derivative
$\dd$, and use (\ref{dop1}) with fixed $\epsilon$. The resulting
expressions are:
\be e=e^t
\left[\dd\!+\!\frac{1\!+\!\varepsilon\!+\!\xi}{2}\right]_q[\omega+\dd]_q
e^{\frac{\pi i\tau}{{\mathcal {N}}^2} (2\dd+1)},\;\;\;
f=e^{-\frac{\pi i\tau}{{\mathcal {N}}^2}
(2\dd+1)}e^{-t}\frac{\left[\frac{\xi-\varepsilon\!+\!1}{2}\!-\!\dd\right]_q}{
[\omega+\dd-1]_q},\;\;\; k=q^{\varepsilon+2\dd}.
 \ee

\section{$R$-matrices, cyclic representations and Yang-Baxter
equations
}
 \addtocounter{section}{0}\setcounter{equation}{0}

One can verify by straight construction that at $q^4=1$ and
$\mathrm{x}\neq 0,\; \mathrm{y}\neq 0$ there are two possible
types of  non-reducible representations, which are $2$-dimensional
cyclic irreps and $4$-dimensional indecomposable
representations (of $A$ or $B$ class) \cite{ak}. 
The tensor product of two general cyclic irreps  usually
decomposes into a sum of two another cyclic irreps with  definite
values of the Casimir operators. It
 follows from (\ref{efk}), that the
  parameters $\mathrm{x},\;\mathrm{y},\;\mathrm{z}$ are the same for two cyclic irreps
 arisen in the fusion. 
Indecomposable representation can appear for some special cases,
under the necessary (but not sufficient) condition that the values
of the quadratic Casimir operator are coinciding.

 Let $R_{ij}$ is an intertwiner matrix of the quantum algebra $sl_q(2)$
defined on the space $V_{i}\otimes V_{j}$, when $V_{i}$ and
$V_{j}$ are cyclic irreps. Hereafter  we shall denote  by index
$i$ in $V_i$ the characteristic index of the representation space
(and not the dimension, as it was in the previous discussion), now
fixing the dimension of the irreps
as $r=2$.  
   As it is known the intertwiner matrices $R_{ij}$, $\check{R}_{ij}=P_{ij}R_{ij}$
 (defined on $V_i\otimes V_j$) which have the
commutativity properties
\bea \Delta R_{ij}=R_{ij}\bar{\Delta},\quad
\Delta\check{R}_{ij}=\check{R}_{ij}\Delta, \label{cop}\eea
%
   satisfy to the Yang-Baxter eqautions \cite{grs,krs,j}
\be R_{ij}R_{ik}R_{jk}=R_{jk}R_{ik}R_{ij} \quad\mbox{or}\quad
\check{R}_{ij}\check{R}_{jk}\check{R}_{ij}=\check{R}_{jk}\check{R}_{ij}\check{R}_{jk}.\label{rrr}
\ee
The spectral parameter dependent YBE, when $R_{ij}$ depends on
$\mathbb{C}$-valued spectral parameters $u_i,\; u_j$, can be
achieved by the affine extension of the quantum algebra (or by any
so-called "baxterization" procedure). In the present situation the
parameters arise naturally connected with the characteristic
parameters of the representations $V_i,\;V_j$ . When the operators
in the l.h.s. and r.h.s. of the YBE act on the tensor product
$V_1\otimes V_2\otimes V_3$, where $V_i$ ($i=1,2,3$) are
characterized by the parameters
$\{\mathrm{x}_i,\mathrm{y}_i,\mathrm{z}_i\}$, then we can take the
YB equations in the following  form (the YBE here are
inhomogeneous in the sense that the representation spaces $V_i,\;
V_j$ on which $\check{R}_{ij}$-matrix acts in general have
different characteristics)
\be
R_{12}({}^{\mathrm{x}_1,\mathrm{y}_1,\mathrm{z}_1}_{\mathrm{x}_2,\mathrm{y}_2,\mathrm{z}_2})
R_{13}({}^{\mathrm{x}_1,\mathrm{y}_1,\mathrm{z}_1}_{\mathrm{x}_3,\mathrm{y}_3,\mathrm{z}_3})R_{23}({}^{\mathrm{x}_2,\mathrm{y}_2,\mathrm{z}_2}_{\mathrm{x}_3,\mathrm{y}_3,\mathrm{z}_3})
=R_{23}({}^{\mathrm{x}_2,\mathrm{y}_2,\mathrm{z}_2}_{\mathrm{x}_3,\mathrm{y}_3,\mathrm{z}_3})
R_{13}({}^{\mathrm{x}_1,\mathrm{y}_1,\mathrm{z}_1}_{\mathrm{x}_3,\mathrm{y}_3,\mathrm{z}_3})R_{12}({}^{\mathrm{x}_1,\mathrm{y}_1,\mathrm{z}_1}_{\mathrm{x}_2,\mathrm{y}_2,\mathrm{z}_2}),\label{rrrx}
\ee
%

 When
$\mathcal{N}=2$, 
we define  two dimensional irreps $V_i$ so, that the algebra
generators have the following general matrix representations on
it:
\bea
e_{i}=\left(\ba{cc}0&\mathrm{x}^a_i\\\frac{\mathrm{x}_i}{\mathrm{x}^a_i}&0\ea\right),\quad
f_i=\left(\ba{cc}0&\frac{\mathrm{y}_i}{\mathrm{y}^a_i}\\\mathrm{y}^a_i&0\ea\right),\quad
k_i=e^{\varepsilon_i}\left(\ba{cc} i&0\\0&-i\ea\right).\eea
Here the algebra relations imply $2 \mathrm{y}^a_i
\mathrm{x}^a_i=\cosh{\varepsilon_i}\mp\sqrt{4 \mathrm{x}_i
\mathrm{y}_i+\left(\cosh{\varepsilon_i}\right)^2}$. So
$e_i^2=\mathrm{x}_i \mathbb{I}$, $f_i^2=\mathrm{y}_i \mathbb{I}$,
$k_i^2=-e^{2 \varepsilon_i} \mathbb{I}$ (we set
$\mathrm{z}_i=-e^{2\varepsilon_i}$)
 and $\mathrm{c}_i=\mp\sqrt{ \mathrm{x}_i
\mathrm{y}_i+\left(\cosh{\varepsilon_i}\right)^2/4} \;\mathbb{I}$,
where $\mathbb{I}$ is the unit operator. In the further discussion
instead of the parameters
$\mathrm{x}_i,\;\mathrm{y}_i,\;\mathrm{z}_i$ we are fixing the
parameters $\mathrm{x}_i,\;\mathrm{c}_i,\;\varepsilon_i$. The
value of $\mathrm{y}^a_i$, using the above relation, can be
written as $ \mathrm{y}^a_i=\frac{1}{\mathrm{x}^a_i}(\frac{\cosh{
\varepsilon_i}}{2}+\mathrm{c}_i)$. Then the parameter
$\mathrm{x}_i^a$ is just a parameter connected with the
automorphism of the algebra: it can be cancelled by the
 automorphism: $g_i\to U_i g_i U_i^{-1}$, $g=e,\; f,\;k$, and
$U=\left(^{\sqrt{\mathrm{x}_i^a}\;\;0}_{0\;\;\;\frac{1}{\sqrt{\mathrm{x}_i^a}}}\right)$.
 However for more generality we take the matrices dependent over the
 parameters $\mathrm{x}_{i,j}^a$.

The generators from the center of the algebra $\breve{c}=e^2,\;
f^2, \; k^2,\; c$ are proportional to the identity operator on the
irreps, and the relation (\ref{cop}) means,  that an intertwiner
can exist only on the such vector spaces' products $V_i\otimes
V_j$, on which, particularly,
$\Delta_{ij}[\breve{c}]=\Delta_{ji}[\breve{c}]$. This means, as it
was stated in the Section 2 for general values of  $\mathcal{N}$
and as we can verify by straight derivation, that the following
relations must be fulfilled:
\bea \mathrm{x}_j (1+e^{2\varepsilon_i})=\mathrm{x}_i
(1+e^{2\varepsilon_j}),\;\;\; \mathrm{c}_j \cosh{\varepsilon_i}=
\pm \mathrm{c}_i \cosh{\varepsilon_j},\label{xce}\eea
where instead of the parameter $\mathrm{y}$ in (\ref{xyz}) we use
the eigenvalues of the quadratic Casimir operator.

Summarizing, we see that the intertwiner matrices $R_{ij}$ for the
general cyclic irreps depend on the  representation
characteristics $\varepsilon_i,\;  \mathrm{x}^a_i$ and
$\varepsilon_j,\; \mathrm{x}^a_j$, as the remaining parameters
$\mathrm{x}_i,\; \mathrm{x}_j,\; \mathrm{c}_i,\;\mathrm{c}_j$ can
be obtained from the relations (\ref{xce}), introducing
appropriate constants
$\mathrm{x}_i/(1+e^{2\varepsilon_i})=\mathrm{x}_0,\;\;\;
\mathrm{c}_i/\cosh{\varepsilon_i}=\mathrm{c}_0$. The parameters
$\mathrm{x}_0$ and $\mathrm{c}_0$ are the same for the all three
$\check{R}$-matrices, so these are constant parameters and can not
be considered as spectral parameters. Let $ \mathrm{c}_j
\cosh{\varepsilon_i}=\mathrm{c}_i \cosh{\varepsilon_j}$, then the
YB equations
 can be presented as:
 \bea&
R_{12}(u_1,u_2;{}^{\varepsilon_1,\mathrm{x}^a_1}_{\varepsilon_2,\mathrm{x}^a_2})
R_{13}(u_1,u_3;{}^{\varepsilon_1,\mathrm{x}^a_1}_{\varepsilon_3,\mathrm{x}^a_3})R_{23}
(u_2,u_3;{}^{\varepsilon_2,\mathrm{x}^a_2}_{\varepsilon_3,\mathrm{x}^a_3})
=&\\\nn
&R_{23}(u_2,u_3;{}^{\varepsilon_2,\mathrm{x}^a_2}_{\varepsilon_3,\mathrm{x}^a_3})
R_{13}(u_1,u_3;{}^{\varepsilon_1,\mathrm{x}^a_1}_{\varepsilon_3,\mathrm{x}^a_3})
R_{12}(u_1,u_2;{}^{\varepsilon_1,\mathrm{x}^a_1}_{\varepsilon_2
,\mathrm{x}^a_2}),& \label{rrrce} \eea
Here for more generality  we introduced additional spectral
parameters $u_i$. However we shall see that it is not necessary to
separate these parameters, they appear naturally.


As in our previous works \cite{shkh,SHKKH}, here we shall look for
the YBE solutions in the form of linear composition of the
invariant operators - projectors. We 
consider as projector operators a definite basis (linearly
independent and complete set)  in the space of the algebra
invariant operators which are commutative with the algebra
generators in the given representation space. Let the last
consists of the irreps $V_i$, which have different
characteristics.  Then the projectors $P_i$ are defined as the
matrices which act on the irreps $V_i$ as unity matrices and
vanish on the another irreps: $P_i\cdot V_j=\delta_{ij} V_j$. When
there are irreps $V_i$, $i=1,...,p$, with the same characteristics
then there are also the projectors $P_{ij}\cdot V_k=P_{jk}V_i$.
The projectors satisfy the following relations:
\bea \sum_i P_i=\mathbb{I},\;\;\; P_i P_j=\delta_{ij}P_i,\;\;\;
P_k P_{ij}=\delta_{ki}P_{ij},\;\;\;P_{ij}P_{kr}=\delta_{jk}P_{ir}.
\eea

In general the tensor product
 $V_{i}\otimes V_{j}$ decomposes into two cyclic irreps, on
 which the Casimir operators $e^2,\; f^2$ and $k^2$ have the same values
 (on the tensor product they
 act as the operators proportional to unity matrix),
 and the Casimir operator $c$ has two different values $\mathrm{c}_{ij},\;
  \bar{c}_{ij}$, differing by a
 sign $\mathrm{c}_{ij}=- \bar{c}_{ij}=-i\mathrm{c}_i 
\sinh{[\varepsilon_i+\varepsilon_j]}/ \cosh{\varepsilon_i}$.
Taking into account this, we can denote the spaces in the tensor
expansion as $V^{\pm}_{ij}$: $V_{i}\otimes V_{j}=V_{ij}^{+}\oplus
V_{ij}^{-}$.

 As we intend to investigate
 the $B$-type representations step by step, here we do not consider
 the indecomposable representations. It
 is worthy to mention however that for the cases described by $\mathrm{c}_i
 \cosh{\varepsilon_j}=\pm \mathrm{c}_j \cosh{\varepsilon_i}
 $, the tensor product $V_2 \otimes V_2=V_2\oplus
 V_2$  under the condition $e^{2(\varepsilon_i+
 \varepsilon_j)}=1$ deforms into $V_2 \otimes V_2=\mathcal{I}^{(4)}_{3,1}$ \cite{kkh}, which
 is an $A$-type indecomposable representation. Now the $\check{R}$-matrix,
 (as well as any invariant matrix) decomposes into the sum of the
 projectors $P_{\mathcal{I}}$ and $P'_{\mathcal{I}}$ (see for  the
 description the work \cite{shkh}). As the number of projection
 operators
 does not increase, the new projectors can be found as the limit cases
 of the linear combinations of the non-deformed
  projectors $P_{\pm}$, and as
 a result no new solutions to YBE arise \cite{shkh}, all the solutions can
 be obtained from the presented solutions taking a proper limit
 $\varepsilon_j\to -
 \varepsilon_i$. When $\mathrm{c}_i=\mathrm{c}_j=0$ the
 tensor product remains the same.

\section{Solutions to YBE}

\addtocounter{section}{0}\setcounter{equation}{0}

 We  analyze in this section the solutions to YBE defined on the tensor product of
 three two-dimensional cyclic irreps. Semi-cyclc and nilpotent
 cases can be obtained taking the particular limits.

  Below we consider separately three different cases corresponding
to the relations (\ref{xce}): $\mathrm{c}_j
\cosh{\varepsilon_i}=\mathrm{c}_i \cosh{\varepsilon_j}$,
$\mathrm{c}_j \cosh{\varepsilon_i}=-\mathrm{c}_i
\cosh{\varepsilon_j}$ and $\mathrm{c}_j =\mathrm{c}_i =0$. The
case $ \cosh{\varepsilon_i}= \cosh{\varepsilon_j}=0$, which occurs
to be degenerated, also will be considered.

\subsection{$\mathrm{c}_j
\cosh{\varepsilon_i}=\mathrm{c}_i \cosh{\varepsilon_j}$}
At first let us explore the case $\mathrm{c}_j
\cosh{\varepsilon_i}=\mathrm{c}_i \cosh{\varepsilon_j}$. There are
two projectors here
 $P_{+}=-(c-\mathrm{c}_{ij}\mathbb{I})/(2\mathrm{c}_{ij})$ and
 $P_{-}=(c+\mathrm{c}_{ij}\mathbb{I})/(2\mathrm{c}_{ij})$: 
  $P_{\pm}\cdot V_{ij}^{\pm}=V_{ij}^{\pm}$. The commutativity relation (\ref{cop}) means that $\check{R}_{ij}$
is  a sum over the "projectors" $\breve{P}_{\pm}=\mathcal{P}_{ij}
P_{\pm}$, where $\mathcal{P}_{ij}$ is an identical transformation
map $V\{x_i,y_i,z_i\}\otimes V\{x_j,y_j,z_j\}\to
V\{x_i,y_i,z_i\}\otimes V\{x_j,y_j,z_j\}$.

The operator $\mathcal{P}_{ij}$ depends for the discussed case on
 the parameters $
 \varepsilon_i,\; \varepsilon_j,\;(\mathrm{x}_j^a,\;\mathrm{x}_i^a)$,
\bea \label{pri}\mathcal{P}_{ij}=\left(\ba{cccc}1&0&0&0\\
0&\frac{\mathrm{x}_j^a}{\mathrm{x}_i^a}\frac{1+e^{2\varepsilon_i}}{1+e^{\varepsilon_i}
 e^{\varepsilon_j}}&\frac{i(e^{\varepsilon_j}-e^{\varepsilon_i})}{1+e^{\varepsilon_i}
 e^{\varepsilon_j}}&0\\
0&\frac{i(e^{\varepsilon_i}-e^{\varepsilon_j})}{1+e^{\varepsilon_i}
 e^{\varepsilon_j}}&\frac{\mathrm{x}_i^a}{\mathrm{x}_j^a}\frac{1+e^{2\varepsilon_j}}{1+e^{\varepsilon_i}
 e^{\varepsilon_j}}&0
\\0&0&0&1\ea\right). \eea
%
This projector operator has the following properties,
$\mathcal{P}_{ij}\mathcal{P}_{ji}=\mathbb{I}$ and
$\mathcal{P}_{ii}=\mathbb{I}$.

 The matrix
$\check{R}_{ij}^{+}(u)=\breve{P}_{+}+f_{ij}\breve{P}_{-}$,
  \bea
\check{R}_{ij}^{+}= \left(\ba{cccc} 1 &0&0&0\\
0&\frac{
\mathrm{x}^a_j(e^{\varepsilon_i}-e^{-\varepsilon_j}f_{ij})\cosh{[\varepsilon_i]}
}{\mathrm{x}^a_i\sinh{[\varepsilon_i+\varepsilon_j]}
}&\frac{i(f_{ij}\cosh{[\varepsilon_i]}-\cosh{[\varepsilon_j]})}
{\sinh{[\varepsilon_i+\varepsilon_j]}}&0\\
0&\frac{i(f_{ij}\cosh{[\varepsilon_j]}-\cosh{[\varepsilon_i]})}
{\sinh{[\varepsilon_i+\varepsilon_j]}}&\frac{
\mathrm{x}^a_i(e^{\varepsilon_j}f_{ij}-e^{-\varepsilon_i})\cosh{[\varepsilon_j]}
}{\mathrm{x}^a_j\sinh{[\varepsilon_i+\varepsilon_j]}
}&0\\
0&0&0& f_{ij}\ea\right), \eea admits a 
 general
solution
  with $f_{ij}=
(f_i
+e^{\varepsilon_i+\varepsilon_j}f_j)/(e^{\varepsilon_i+\varepsilon_j}
f_i+f_j)$.  Here the
 coefficients $f_i, f_j$ are arbitrary, and enter into the solution as $f_i/f_j$, so we
 can denote that proportion as
  $\frac{f_i}{f_j}\equiv\frac{f(\varepsilon_i,\mathrm{x}^a_i,\{u_i\})}{f(\varepsilon_j,\mathrm{x}^a_j,\{u_j\})}$
   with arbitrary function $f(\varepsilon_i,\mathrm{x}_i^a,\{u_i\})$ and
 a set of the spectral parameters $\{u_i,\; u_j\}$.
 The corresponding matrix is
\bea
\check{R}_{ij}^+(u,e^{\varepsilon_i},e^{\varepsilon_j},\mathrm{x}_i^a,\mathrm{x}_j^a)=
\left(\ba{cccc}1&0&0&0\\
0&\frac{\mathrm{x}^a_j}{\mathrm{x}^a_i}\frac{(1+e^{2\varepsilon_i})\frac{f_i}{f_j}}{1+e^{\varepsilon_i+\varepsilon_j}
\frac{f_i}{f_j}}&
i\frac{e^{\varepsilon_i}-e^{\varepsilon_j} \frac{f_i}{f_j}}{1+e^{\varepsilon_i+\varepsilon_j} \frac{f_i}{f_j}}&0\\
0&i\frac{e^{\varepsilon_j}-e^{\varepsilon_i}
\frac{f_i}{f_j}}{1+e^{\varepsilon_i+\varepsilon_j}
\frac{f_i}{f_j}}&
\frac{\mathrm{x}^a_i}{\mathrm{x}^a_j}\frac{(1+e^{2\varepsilon_j})}{1+e^{\varepsilon_i+\varepsilon_j} \frac{f_i}{f_j}}&0\\
0&0&0&\frac{\frac{f_i}{f_j}+e^{\varepsilon_i+\varepsilon_j}}{1+e^{\varepsilon_i+\varepsilon_j}
\frac{f_i}{f_j}}\ea\right). \label{rzij} \eea

 Note, that the matrix 
(\ref{rzij})
for the particular homogeneous case
 $\varepsilon_i=\varepsilon_j,\;
\mathrm{x}_i^a=\mathrm{x}_j^a$, is
a  $\check{R}$-matrix, describing the $XX$-model in the transverse
magnetic field ($\cos{\varepsilon}$). Setting
$f_i/f_j=e^{u_i-u_j}\equiv e^u$, and
%
 %
 %
after consecutive replacements $e^{2\varepsilon}=e^{2u_0}$, $u\to
i u$ and $u_0\to i u_0+i \pi/2$ and multiplying the matrix  by an
overall function $\sin(u+u_0)$, we shall come to the
$\check{R}$-matrix, which describes the $XX$-model in the
transverse field $\cos{u_0}$ \cite{shs,grs,DA}.
\bea
\check{R}_{ij}(u)=\left(\ba{cccc}\sin(u+u_0)&0&0&0\\
0&e^{i u}\sin{u_0}&\sin{u}&0\\
0&\sin{u}&e^{-i u}\sin{u_0}&0
\\0&0&0&\sin(u_0-u)\ea\right).\label{ru0}\eea
This matrix satisfies to the simple YBE
\be
R_{12}(u-v)R_{13}(u)R_{23}(v)=R_{23}(v)R_{13}(u)R_{12}(u-v).\label{rrrsu}
\ee

As there are classified the YBE solutions with general
non-homogeneous $4\times 4$  $R$-matrices with eight non-zero
matrix elements (like eight-vertex model's $R$-matrix
\cite{rbaxter}) depending on the one spectral parameter
("difference property": $R(u,v)=R(u-v)$) \cite{rbaxter,KS}, we
know that there are limited kind of such solutions and all the
interesting cases are restricted with the cases of the
$XYZ$-model's matrix and the "free-fermionic" non-homogeneous
extensions, one of which is just the matrix brought above. At
$q=i$ the $sl_q(2)$ invariant matrices defined on the irreps all
have  free-fermionic property:
$\check{R}_{00}^{00}\check{R}_{11}^{11}=\check{R}_{01}^{01}\check{R}_{10}^{10}-
\check{R}_{01}^{10}\check{R}_{10}^{01}$ (see the Summary).

%

At the end of this subsection we want to mention the relation of
the solution (\ref{rzij}) to the one obtained in the paper
\cite{RGS}. These two solutions can be related by an automorphism
of the matrix $R_{ij}$, written as
\bea R_{n_i n_j}^{p_i p_j}\Rightarrow R_{n_i n_j}^{p_i
p_j}\frac{\mathrm{f}_{n_i}\mathrm{f}_{n_j}}{\mathrm{f}_{p_i}\mathrm{f}_{p_j}},
\label{auf} \eea
induced from the transformations $\mathrm{e}_{n_{i,j}}\to
\mathrm{f}_{n_{i,j}}\mathrm{e}_{n_{i,j}}$  of the vector basis
$\mathrm{e}_{n_{i,j}}$ ($n_{i,j}=0,1$, $p_{i,j}=0,1$) of the space
$V_{i,j}$  with a function $\mathrm{f}_{n_{i,j}}=1+i
e^{\varepsilon_{i,j}}\delta_{n_{i,j},1}$. So, only the matrix
elements $\check{R}_{01}^{01}$ and $\check{R}_{10}^{10}$ transform
correspondingly into the functions
$\frac{\mathrm{x}^a_j}{\mathrm{x}^a_i}\frac{(1-i
e^{\varepsilon_i})(1+i
e^{\varepsilon_j})\frac{f_i}{f_j}}{1+e^{\varepsilon_i+\varepsilon_j}
\frac{f_i}{f_j}}$ and
$\frac{\mathrm{x}^a_i}{\mathrm{x}^a_j}\frac{(1-i
e^{\varepsilon_j})(1+i
e^{\varepsilon_i})}{1+e^{\varepsilon_i+\varepsilon_j}
\frac{f_i}{f_j}}$, being equivalent to the matrix elements in
\cite{RGS}.

Now let us represent the next solutions to the YBE with the cyclic
representations.
\subsection{$\mathrm{c}_j
\cosh{\varepsilon_i}=-\mathrm{c}_i \cosh{\varepsilon_j}$}
 In the case $\mathrm{c}_j
\cosh{\varepsilon_i}=-\mathrm{c}_i \cosh{\varepsilon_j}$ the
transformation operator $\mathcal{P}_{ij}$ has the following
matrix representation
\bea \label{pi}\mathcal{P}_{ij}=\left(\ba{cccc}
\frac{i(e^{\varepsilon_i}+e^{\varepsilon_j})}{1-e^{\varepsilon_i+\varepsilon_j}}
&0&0&\frac{\mathrm{x}_i(1+e^{2\varepsilon_j})}{(1-e^{\varepsilon_i+\varepsilon_j})\mathrm{x}^a_i
\mathrm{x}^a_j}\\
0&0&1&0\\
0&1&0&0\\
\frac{\mathrm{x}^a_i
\mathrm{x}^a_j(1+e^{2\varepsilon_i})}{\mathrm{x}_i(1-e^{\varepsilon_i+\varepsilon_j})}&0&0&
\frac{i(e^{\varepsilon_i}+e^{\varepsilon_j})}{-1+e^{\varepsilon_i+\varepsilon_j}}\ea\right).\eea
As we see there is a dependence from the parameter $\mathrm{x}_i$.
Recalling, that $\mathrm{x}_i/(1+e^{2
\varepsilon_i})=\mathrm{x}_j/(1+e^{2 \varepsilon_j})$, we can use
an independent parameter $\mathrm{x}_0=\mathrm{x}_i/(1+e^{2
\varepsilon_i})$ instead of $\mathrm{x}_i$. And then the matrix
$\check{R}_{ij}^-=\mathcal{P}_{ij}(P_++g_{ij}P_-)$  is
 \bea
\check{R}_{ij}^-=\left(\ba{cccc}
\frac{-i(g_{ij}\cosh{[\varepsilon_i]}+\cosh{[\varepsilon_j]})}{\sinh{[\varepsilon_i+\varepsilon_j]}}
&0&0&\frac{-2
\mathrm{x}_0(g_{ij}+e^{\varepsilon_i+\varepsilon_j})\cosh{[\varepsilon_i]}\cosh{[\varepsilon_j]}}{\sinh{[\varepsilon_i+\varepsilon_j]}\mathrm{x}^a_i
\mathrm{x}^a_j}\\
0&0&1&0\\
0&g_{ij}&0&0\\
\frac{- \mathrm{x}^a_i
\mathrm{x}^a_j(g_{ij}+e^{-\varepsilon_i-\varepsilon_j})}{2
\mathrm{x}_0\sinh{[\varepsilon_i+\varepsilon_j]}}&0&0&
\frac{i(\cosh{[\varepsilon_i]}+g_{ij}\cosh{[\varepsilon_j]})}{\sinh{[\varepsilon_i+\varepsilon_j]}}\ea\right).
\eea
The matrix of this kind 
have to be considered in the YBE together with the matrix
$\check{R}_{ij}^+=\mathcal{P}_{ij}(P_++f_{ij}P_-)$ in the
following non-homogeneous YBE,
  \bea&
R_{12}^+(u_1,u_2;{}^{\varepsilon_1,c_1,\mathrm{x}^a_1}_{\varepsilon_2,c_2,\mathrm{x}^a_2})
R_{13}^-(u_1,u_3;{}^{\varepsilon_1,c_1,\mathrm{x}^a_1}_{\varepsilon_3,c_3,\mathrm{x}^a_3})R_{23}^-
(u_2,u_3;{}^{\varepsilon_2,c_2,\mathrm{x}^a_2}_{\varepsilon_3,c_3,\mathrm{x}^a_3})
=&\\\nn
&R_{23}^-(u_2,u_3;{}^{\varepsilon_2,c_2,\mathrm{x}^a_2}_{\varepsilon_3,c_3,\mathrm{x}^a_3})
R_{13}^-(u_1,u_3;{}^{\varepsilon_1,c_1,\mathrm{x}^a_1}_{\varepsilon_3,c_3,\mathrm{x}^a_3})
R_{12}^+(u_1,u_2;{}^{\varepsilon_1,c_1,\mathrm{x}^a_1}_{\varepsilon_2,c_2,\mathrm{x}^a_2}),&\label{rrrw}
\eea
where the
 conditions
 $c_1
\cosh{\varepsilon_2}=c_2 \cosh{\varepsilon_1}$, $c_1
\cosh{\varepsilon_3}=-c_3 \cosh{\varepsilon_1}$ and $c_2
\cosh{\varepsilon_3}=-c_3 \cosh{\varepsilon_2}\;\;\;$
 work.

 The solutions to the presented YBE are of this graceful form
\bea f_{ij}=
\frac{f[u_i,\varepsilon_i,\mathrm{x}^a_i]+e^{\varepsilon_i+\varepsilon_j}f[u_j,\varepsilon_j,\mathrm{x}^a_j]}
{f[u_i,\varepsilon_i,\mathrm{x}^a_i]e^{\varepsilon_i+\varepsilon_j}+f[u_j,\varepsilon_j,\mathrm{x}^a_j]},\;\;\;
 g_{ij}=
\frac{f[u_i,\varepsilon_i,\mathrm{x}^a_i]-e^{\varepsilon_i+\varepsilon_j}g[u_j,\varepsilon_j,\mathrm{x}^a_j]}
{-f[u_i,\varepsilon_i,\mathrm{x}^a_i]e^{\varepsilon_i+\varepsilon_j}+g[u_j,\varepsilon_j,\mathrm{x}^a_j]},
\eea
where the functions $f[u,\varepsilon,\mathrm{x}^a]$,
$g[u,\varepsilon,\mathrm{x}^a]$ are arbitrary. Note, that the
solution $\check{R}_{ij}^+$ coincides with the general solution
obtained in the previous subsection.

{\it{Note}}, that the resemblance of the functions $f_{ij}$ and
$g_{ij}$ is not casual, as the constraint $\mathrm{c}_j
\cosh{\varepsilon_i}=-\mathrm{c}_i \cosh{\varepsilon_j}$ can be
transformed into  $\mathrm{c}_j \cosh{\varepsilon_i}=\mathrm{c}_i
\cosh{(\varepsilon_j+i\pi)}$ (corresponding to the case discussed
in the previous subsection), which means that we can consider the
space $V_j$ having parameter $(\varepsilon_j+i\pi)$ instead of
$\varepsilon_j$, which does not change the values of
$\mathrm{z}_j,\;\mathrm{x}_j,\;\mathrm{y}_j$, but interchanges the
vector states: $\{v_1,\;v_2\}_j\to \{v_2,\;v_1\}_j$, explaining
thus the difference between the matrix forms of (\ref{pi}) and
(\ref{pri}).  And moreover, we can extend this observation for the
case with general $\mathcal{N}$.  Then the relations between the
characteristics of two cyclic irreps $V_i,\; V_j$, on which an
intertwiner is defined can be presented as follows from the
general constraints (\ref{constr}) and (\ref{xyz})
($q^{\mathcal{N}}=\pm 1$):
\bea
\frac{x_i}{\left(\mathrm{z}_i^{1/2}-\mathrm{z}_i^{-1/2}\right)^2}=
\frac{x_j}{\left(\mathrm{z}_j^{1/2}-\mathrm{z}_j^{-1/2}\right)^2},\quad
\frac{q^{\mathcal{N} \xi_i/2}+(\mp
1)^{\mathcal{N}}q^{-\mathcal{N}\xi_i/2}}{\mathrm{z}_i^{1/2}-\mathrm{z}_i^{-1/2}}
=\pm \frac{q^{\mathcal{N} \xi_j/2}+(\mp
1)^{\mathcal{N}}q^{-\mathcal{N}\xi_j/2}}{\mathrm{z}_j^{1/2}-\mathrm{z}_j^{-1/2}}.
\eea
The second equations connected with the quadratic Casimir
operators with two signs can be relate one to another by the
change $z_j^{1/2}\to -z_j^{1/2}$.

\subsection{$\mathrm{c}_i=\mathrm{c}_j=0$}
The next case corresponds to the situation, when
$\mathrm{c}_i=\mathrm{c}_j=0$.
  Now two eigenvalues of the Casimir operator $\mathrm{c}_{ij}$ coincide one  with another
  and equal to $0$. It means that there are four linear independent projection operators,
   which compose the $\check{R}$-matrix. We denote them as
    $\mathcal{P}_{ij}\cdot\{P_{++},\; P_{--},\; P_{+-},\; P_{-+}\}$. The first two operators
    act on the each of  two cyclic representations as identity operator and
    vanish on the other irrep ($P_{\pm\pm}\cdot V_{ij}^{\pm}=V_{ij}^{\pm}$,
    $P_{\pm\pm}\cdot V_{ij}^{\mp}=0$), meanwhile two other projectors transpose one
     irrep with the other ($P_{\pm\mp}\cdot V_{ij}^{\mp}=V_{ij}^{\pm}$, $P_{\pm\mp}\cdot V_{ij}^{\pm}=0$).
      The transformation operator $\mathcal{P}_{ij}$ now can be written as
%
\bea \mathcal{P}_{ij}=\left(\ba{cccc} \frac{(\mathrm{x}^a_i)^2
e^{\varepsilon_j}\cosh{[\varepsilon_j]}-e^{-\varepsilon_j}(\mathrm{x}^a_i)^2\cosh{[\varepsilon_i]}}{\mathrm{x}^a_i
\mathrm{x}^a_j \sinh{[\varepsilon_i+\varepsilon_j]}} &0&0&\frac{2i
\mathrm{x}_0\left(
e^{\varepsilon_i}(\cosh{[\varepsilon_i]}/\mathrm{x}^a_i)^2-e^{\varepsilon_j}(\cosh{[\varepsilon_j]}/\mathrm{x}^a_j)^2\right)}{
 \sinh{[\varepsilon_i+\varepsilon_j]}}\\
0&0&1&0\\
0&1&0&0\\
\frac{(\mathrm{x}^a_i)^2 e^{-\varepsilon_i}-(\mathrm{x}^a_j)^2
e^{-\varepsilon_j}}{2 i
\mathrm{x}_0\sinh{[\varepsilon_i+\varepsilon_j]}}&0&0&
\frac{(\mathrm{x}^a_j)^2
e^{\varepsilon_i}\cosh{[\varepsilon_i]}-e^{-\varepsilon_i}(\mathrm{x}^a_i)^2\cosh{[\varepsilon_j]}}{\mathrm{x}^a_i
\mathrm{x}^a_j
\sinh{[\varepsilon_i+\varepsilon_j]}}\ea\right).\eea

Then the matrix $\check{R}_{ij}=
\mathcal{P}_{ij}\left(P_{++}+f_{ij} P_{--}+g_{ij}P_{+-}+h_{ij}
P_{-+}\right)$ has the following form
\bea
&{\small{\check{R}_{ij}\!=\!\left(\!\!\!\ba{cccc}\frac{\mathrm{x}^a_i
e^{\varepsilon_j}\cosh{[\varepsilon_j]}}{
\mathrm{x}^a_j\sinh{[\varepsilon_i+\varepsilon_j]}}&0&0&\frac{2
\mathrm{x}_0 e^{\varepsilon_j}(\cosh{[\varepsilon_j]})^2}{i
(\mathrm{x}^a_j)^2\sinh{[\varepsilon_i+\varepsilon_j]}}\\
0&1&0&0\\
0&0&0&0\\
\frac{(\mathrm{x}^a_i)^2 e^{-\varepsilon_i}}{2 i
\mathrm{x}_0\sinh{[\varepsilon_i+\varepsilon_j]}}&0&0&\frac{-\mathrm{x}^a_i
e^{-\varepsilon_i}\cosh{[\varepsilon_j]}}{
\mathrm{x}^a_j\sinh{[\varepsilon_i+\varepsilon_j]}}\ea\!\!\!\right)\!+\!f_{ij}\!
\left(\!\!\!\ba{cccc}\frac{-\mathrm{x}^a_j
e^{-\varepsilon_j}\cosh{[\varepsilon_i]}}{
\mathrm{x}^a_i\sinh{[\varepsilon_i+\varepsilon_j]}}&0&0&\frac{2 i
\mathrm{x}_0 e^{\varepsilon_i}(\cosh{[\varepsilon_i]})^2}{
(\mathrm{x}^a_i)^2\sinh{[\varepsilon_i+\varepsilon_j]}}\\
0&0&0&0\\
0&0&1&0\\
\frac{i(\mathrm{x}^a_j)^2 e^{-\varepsilon_j}}{2
\mathrm{x}_0\sinh{[\varepsilon_i+\varepsilon_j]}}&0&0&\frac{\mathrm{x}^a_j
e^{\varepsilon_i}\cosh{[\varepsilon_i]}}{
\mathrm{x}^a_i\sinh{[\varepsilon_i+\varepsilon_j]}}\ea\!\!\!\!\!\right)}}&\label{rco}\\&{
\small{+ \!g_{ij}\!\left(\!\ba{cccc}\frac{
\cosh{\!\![\varepsilon_j]}}{i\sinh{\![\varepsilon_i+\varepsilon_j]}}&0&0&\frac{-2
\mathrm{x}_0
e^{\varepsilon_i+\varepsilon_j}\cosh{\![\varepsilon_j]}\cosh{\![\varepsilon_i]}}{
\mathrm{x}^a_i \mathrm{x}^a_j \sinh{\![\varepsilon_i+\varepsilon_j]}}\\
0&0&1&0\\
0&0&0&0\\
\!\!\!\frac{-\mathrm{x}^a_i \mathrm{x}^a_j
e^{-\varepsilon_i-\varepsilon_j}}{ 2
\mathrm{x}_0\sinh{\![\varepsilon_i+\varepsilon_j]}}&0&0&\frac{i
\cosh{\!\![\varepsilon_i]}}{
\sinh{\![\varepsilon_i+\varepsilon_j]}}\ea\!\!\!\right)\!+\!h_{ij}\!
\left(\!\ba{cccc}\frac{
\cosh{\!\![\varepsilon_i]}}{i\sinh{\![\varepsilon_i+\varepsilon_j]}}&0&0&\frac{-2
\mathrm{x}_0 \cosh{\![\varepsilon_j]}\cosh{\![\varepsilon_i]}}{
\mathrm{x}^a_i \mathrm{x}^a_j \sinh{\![\varepsilon_i+\varepsilon_j]}}\\
0&0&0&0\\
0&1&0&0\\
\!\!\!\frac{-\mathrm{x}^a_i \mathrm{x}^a_j}{ 2
\mathrm{x}_0\sinh{\![\varepsilon_i+\varepsilon_j]}}&0&0&\frac{i
\cosh{\!\![\varepsilon_j]}}{
\sinh{\![\varepsilon_i+\varepsilon_j]}}\ea\!\!\!\right).}}&\nn\eea

Among the YB equations there is simple relation on the coefficient
function $f_{ij}$
\bea f_{ik}=f_{ij}f_{jk}, \eea
which expresses the $factorizable$ property of $f_{ij}$. It means
that we can take $f_{ik}=f_i/f_k$, with the functions $f_a$
($a=i,k$) depending only of the parameters with the index $a$.

At first let us explore two simple cases.

When $g_{ij}=0$ and $h_{ij}=0$, then there is  one solution to YBE
with the following
value of the $factorizable$ function $f_{ij}$ 
%
\be
f_{ij}=\frac{e^{\varepsilon_j}(\cosh{[\varepsilon_j]}\mathrm{x}^a_i)^2
\left(1\pm\sqrt{1+f_0(
\cosh{[\varepsilon_i]})^2}\right)}{e^{\varepsilon_i}(\cosh{[\varepsilon_i]}\mathrm{x}^a_j)^2
\left(1\pm\sqrt{1+f_0(
\cosh{[\varepsilon_j]})^2}\right)},\label{f0}\ee
where $f_0$ is a constant.

When the expression for the $\check{R}$-matrix includes only the
projectors $P_{+-}$ and $P_{-+}$, then there is no solution to the
YBE.

For obtaining the general solutions let us consider at first the
homogeneous solutions which satisfy the conditions
$\mathrm{x}^a_i=\mathrm{x}^a_j,\;\varepsilon_i=\varepsilon_j$, 
 and explore the one-parametric YBE equations (\ref{rrrsu}).

 As we have stated,  the function $f_{ij}$ can be presented as
 $f_{ik}=f_i/f_k$. If $f_i$ depends only on the state parameters
 $\mathrm{x}^a_i$ and $\varepsilon_i$, then in the  homogeneous case
 $f_{ij}=1$.
  There are two such spectral-parameter dependent solutions. One is
  written as
$f_{ij}=1$ and $g_{ij}=-h_{ij}=\tanh{[u]}$ ($u$ is an additive
spectral parameter)
\bea \check{R}^*(u)=\small{\left(\ba{cccc}1&0&0&-e^{\alpha}\tanh{[u]}\\
0&1&\tanh{[u]}&0\\
0&-\tanh{[u]}&1&0\\
e^{-\alpha}\tanh{[u]}&0&0&1 \ea\right),\;\;\;
e^{\alpha}=\frac{2e^{\varepsilon}\cosh{[\varepsilon_i]}\mathrm{x}_0}{(\mathrm{x}^a_i)^2}.}
\label{hg}\eea
The second solution corresponds to $f_{ij}=1$ and
$g_{ij}=h_{ij}=\tanh{[u]}\tanh{[\varepsilon_i]}$,
\bea
\check{R}^{**}(u)=\small{\left(\ba{cccc}1-i\frac{\tanh{[u]}}{\cosh{\varepsilon_i}}&0&0&
-e^{\alpha} \tanh{[u]}\\
0&1&\tanh{[\varepsilon_i]}\tanh{[u]}&0\\
0&\tanh{[\varepsilon_i]}\tanh{[u]}&1&0\\
-e^{-\alpha}\tanh{[u]}&0&0&1+i\frac{\tanh{[u]}}{\cosh{\varepsilon_i}}
\ea\right)}. \label{gh} \eea
 This is just a trigonometric limit of the $\check{R}$-matrix
 of the 2d Ising model \cite{rbaxter,shs,KS}.

 If $f_i$ has also an extra argument $u_i$ (spectral parameter),
then in the  homogeneous case we take $f_i=f[u_i]$, and
$f_{ij}=f[u_i]/f[u_j]$. As we are exploring now one parametric YBE
(\ref{rrrsu}), we require that the function $f_{ij}$ depends on
the difference of the spectral parameters, which dictates the
choice of $f[u_i]$ as an exponential function, and
$f_{ij}=e^{u_i-u_j}\equiv e^u$. Then we shall
come to the solution (\ref{ru0}) obtained in the subsection 4.1. 
 The generalization of this solution to the inhomogeneous case is
\bea \label{f1} f_{ij}=\frac{(\mathrm{x}^a_i)^2}{(\mathrm{x}^a_j)^2}\frac{f[\varepsilon_i,\mathrm{x}^a_i,\{u_i\}]}{f[\varepsilon_j,\mathrm{x}^a_j,\{u_j\}]}\\
\label{gf} g
_{ij}=i\frac{\mathrm{x}^a_i}{\mathrm{x}^a_j}\frac{e^{\varepsilon_i}(1+e^{2\varepsilon_i})\frac{f[\varepsilon_i,\mathrm{x}^a_i,\{u_i\}]}
{f[\varepsilon_j,\mathrm{x}^a_j,\{u_j\}]}-e^{\varepsilon_j}
(1+e^{2\varepsilon_j})}
{(1+e^{2\varepsilon_i})(1+e^{2\varepsilon_j})},\\\label{hf}
h_{ij}=i\frac{\mathrm{x}^a_i}{\mathrm{x}^a_j}
\frac{e^{\varepsilon_j}(1+e^{2\varepsilon_i})\frac{f[\varepsilon_i,\mathrm{x}^a_i,\{u_i\}]}{f[\varepsilon_j,\mathrm{x}^a_j,\{u_j\}]}-e^{\varepsilon_i}(1+e^{2\varepsilon_j})}
{(1+e^{2\varepsilon_i})(1+e^{2\varepsilon_j})}.\eea
The function $f[\varepsilon_i,\mathrm{x}^a_i,\{u_i\}]$ is an
arbitrary function. In the particular homogeneous case when
$\varepsilon_i=\varepsilon_j,\;\mathrm{x}^a_i=\mathrm{x}^a_j$ and
$f_{ij} =e^{2(u_i-u_j)}\equiv e^{2u}$,  we have $f_{ij}=e^{2u},\;
g_{ij}=h_{ij}=i e^{u}\sinh{[u]}/\cosh{[\varepsilon_i]}$, the
corresponding $R$-matrix coincides with the solution (\ref{ru0}).
And one can observe, that in the inhomogeneous case, taking
$\frac{f[\varepsilon_i,\mathrm{x}^a_i,\{u_i\}]}{f[\varepsilon_j,\mathrm{x}_j^a,\{u_j\}]}=
\frac{(1+e^{2\varepsilon_j})f_j}{(1+e^{2\varepsilon_i})f_i}$,
after some normalization calculations this is the solution
$\check{R}^+_{ij}$ (\ref{rzij}) which we have in the subsection
4.1. The appearance of the solution $\check{R}^+_{ij}$ here is not
casual, as the eigenvalues $c_{i,j}$ are not presented in the
projectors evidently, so the values $c_{i,j}=0$ are also
permissible in the case discussed in the subsection 4.1. This
solution, with the choice $\frac{f_i}{f_j} = e^{2u}$ is also
equivalent to the trigonometric limit of the free-fermionic
elliptic solutions \cite{Felder,BSf}, after fixing
the elliptic module as $k=0$.  

 The extension for the first matrix (\ref{hg}) with the parameters
$\mathrm{x}^a_i\neq
\mathrm{x}^a_j,\;\varepsilon_i\neq\varepsilon_j$ can be written as
\bea \label{fa1}f_{ij}=\frac{(\mathrm{x}^a_i)^2}{(\mathrm{x}^a_j)^2}\frac{1+e^{2\varepsilon_j}}{1+e^{2\varepsilon_i}},\\
g_{ij}=-h_{ij}=(1+e^{2\varepsilon_j})\frac{\mathrm{x}^a_i}{\mathrm{x}^a_j}
\frac{\pm(h[\varepsilon_i,\mathrm{x}^a_i,\{u_i\}]-h[\varepsilon_j,\mathrm{x}^a_j,\{u_j\}])}
{h[\varepsilon_i,\mathrm{x}^a_i,\{u_i\}](\pm
i+e^{\varepsilon_i})(e^{\varepsilon_j}\mp i)+
h[\varepsilon_j,\mathrm{x}^a_j,\{u_j\}](\pm
i+e^{\varepsilon_j})(e^{\varepsilon_i}\mp i)}.\label{h1}\eea
The function $h[\varepsilon,\mathrm{x}^a,\{u\}]$, here and below
too, is an arbitrary function. Two solutions with different signs
can be mapped one to another by the shift of the variables
$\varepsilon_{i,j}\to \varepsilon_{i,j}+i\pi$ and transformation
$h_{ij}\to -h_{ij},\; g_{ij}\to -g_{ij}$. The corresponding
$\check{R}$ matrix, after normalization, with multiplication by a
function, has the form (we choose the case with upper sign in
(\ref{h1}) and use the notations
$h_i=h[\varepsilon_i,\mathrm{x}^a_i,\{u_i\}]$,
$h_j=h[\varepsilon_j,\mathrm{x}^a_j,\{u_j\}]$ and
$\bar{h}_i=h_i\frac{e^{\varepsilon_i}+i}{e^{\varepsilon_i}-i}$,
$\bar{h}_j=h_j \frac{e^{\varepsilon_j}+i}{e^{\varepsilon_j}-i}$)
\bea\label{ras}
\check{R}^*_{ij}={\small\left(\ba{cccc}h_i+h_j&0&0&-\frac{x_0(e^{\varepsilon_i}-i)(e^{\varepsilon_j}-i)\left(\bar{h}_i-\bar{h}_j
\right)}{x^a_i
x^a_j}\\
0&\frac{x_j^a(e^{\varepsilon_i}-i)}{x_i^a(e^{\varepsilon_j}+i)}\left(\bar{h}_i+\bar{h}_j\right)&h_i-h_j&0\\
0&h_j-h_i&\frac{x^a_i(e^{\varepsilon_j}-i)}{x^a_j(e^{\varepsilon_i}+i)}\left(\bar{h}_i+\bar{h}_j\right)&0\\
\frac{x^a_i
x^a_j\left(\bar{h}_i-\bar{h}_j\right)}{x_0(e^{\varepsilon_i}+i)(e^{\varepsilon_j}+i)}&0&0&h_i+h_j\ea\right)}
\eea
This matrix, after an appropriate re-parametrization can be
brought to the form of the two-parametric solution of YBE
\cite{KS}, see also (\ref{ruw}).

The extension of the second solution (\ref{gh}) for the
inhomogeneous case is
\bea \label{fa2}f_{ij}=\frac{(\mathrm{x}^a_i)^2}{(\mathrm{x}^a_j)^2}\frac{1+e^{2\varepsilon_j}}{1+e^{2\varepsilon_i}}\\
g_{ij}=h_{ij}+2i\frac{\mathrm{x}^a_i}{\mathrm{x}^a_j}\frac{e^{\varepsilon_i}-e^{\varepsilon_j}}{1+e^{2\varepsilon_i}},\\
h_{ij}=\frac{\mathrm{x}^a_i}{\mathrm{x}^a_j}
\frac{h[\varepsilon_i,\mathrm{x}^a_i,\{u_i\}](1\pm
i(e^{\varepsilon_j}-e^{\varepsilon_i})-e^{\varepsilon_i+\varepsilon_j})-
h[\varepsilon_j,\mathrm{x}^a_j,\{u_j\}](1\pm
i(e^{\varepsilon_i}-e^{\varepsilon_j})-e^{\varepsilon_i+\varepsilon_j})}
{\pm
(1+e^{2\varepsilon_i})(h[\varepsilon_i,\mathrm{x}^a_i,\{u_i\}]+h[\varepsilon_j,\mathrm{x}^a_j,\{u_j\}]
)}.\label{h2}\eea
By redefinition of the arbitrary functions
$h[\varepsilon_i,\mathrm{x}^a_i,\{u_i\}]$, it is possible to
change the appearance of the functions $h_{ij},\; g_{ij}$.
Particularly, one can  bring the parametrization in (\ref{h1}) to
the form $h_{ij}=\frac{\mathrm{x}^a_i}{\mathrm{x}^a_j}
\frac{h[\varepsilon_i,\mathrm{x}^a_i,\{u_i\}](1\pm
i(e^{\varepsilon_j}-e^{\varepsilon_i})+e^{\varepsilon_i+\varepsilon_j})-
h[\varepsilon_j,\mathrm{x}^a_j,\{u_j\}](1\pm
i(e^{\varepsilon_i}-e^{\varepsilon_j})+e^{\varepsilon_i+\varepsilon_j})}
{\pm
(1+e^{2\varepsilon_i})(h[\varepsilon_i,\mathrm{x}^a_i,\{u_i\}]+h[\varepsilon_j,\mathrm{x}^a_j,\{u_j\}]
)}$, similar, but no equal to (\ref{h2}).

The particular homogeneous cases (\ref{hg}, \ref{gh}) correspond
to the choice
$h[\varepsilon_i,\mathrm{x}^a_i,\{u_i\}]/h[\varepsilon_j,\mathrm{x}^a_j,\{u_j\}]=e^{2(u_i-u_j)}\equiv
e^{2u}$. Note, that this solution with the same choice of the
function $h[\varepsilon_i,\mathrm{x}^a_i,\{u_i\}]$, but in
inhomogeneous case $\varepsilon_i\neq \varepsilon_j$ is equivalent
to the trigonometric limit of the elliptic solutions
\cite{Felder,BSf}, with the elliptic module $k=1$ (for the
parameterizations presented in \cite{BSf}, one must perform some
transformations, such as $\varepsilon_i=\varphi_i+\pi/2$ and then
the automorphism (\ref{auf}), with appropriate chosen functions
$\mathrm{f}_{n_i}$).

 The matrix representation of the solutions (\ref{fa2}-\ref{h2})
 is the following (the case with upper sign), where we have used the
 notations $\bar{h}_{ij}=h_i+h_j$, $\tilde{h}_{ij}=h_i-h_j$ and $\varepsilon_{ij}=
 \varepsilon_i+\varepsilon_j$
\bea \label{ras8}\check{R}^{**}_{ij}=\frac{x^a_i}{x^a_j(1+e^{2
\varepsilon_i})}\times \quad\\\nn{\small
{\small\left(\!\!\!\ba{cccc}\frac{(1+e^{\varepsilon_{ij}})
\bar{h}_{ij}-i(e^{\varepsilon_i}+e^{\varepsilon_j})\tilde{h}_{ij}}{
\bar{h}_{ij}}&0&0&
\frac{-(1+e^{2\varepsilon_i})(1+e^{2\varepsilon_j})\tilde{h}_{ij}x_0}{x^a_i x^a_j\bar{h}_{ij}}\\
0&\frac{x^a_j(1+e^{2\varepsilon_i})}{x^a_i}&
\frac{(e^{\varepsilon_{ij}}-1)
\tilde{h}_{ij}+i(e^{\varepsilon_i}-e^{\varepsilon_j})\bar{h}_{ij}}{\bar{h}_{ij}}&0\\
0&\frac{(e^{\varepsilon_{ij}}-1)
\tilde{h}_{ij}-i(e^{\varepsilon_i}-e^{\varepsilon_j})\bar{h}_{ij}}{\bar{h}_{ij}}&\frac{x^a_i(1+e^{2\varepsilon_j})}
{x^a_j}&0\\
\frac{-\tilde{h}_{ij}x^a_i x^a_j}{x_0
\bar{h}_{ij}}&0&0&\frac{(1+e^{\varepsilon_{ij}})
\bar{h}_{ij}+i(e^{\varepsilon_i}+e^{\varepsilon_j})\tilde{h}_{ij}}{\bar{h}_{ij}}\ea\!\!\!\right)}}
\eea

The obtained solutions $\check{R}(\varepsilon_i,\varepsilon_j,
\mathrm{x}^a_i,\mathrm{x}^a_j;u_i,u_j)$ contain arbitrary
functions on the variables $\varepsilon_i,\mathrm{x}^a_i,u_i$.
This dependence from the arbitrary functions has a remarkable
property of "factorization", in the sense, that the functions
appear in the matrix elements only in the form of the ratio
$\frac{f/h[\varepsilon_i,\mathrm{x}^a_i,\{u_i\}]}
{f/h[\varepsilon_j,\mathrm{x}^a_j,\{u_j\}]}$. In this way it gives
us an opportunity to choose the dependence from the extra
arguments (spectral parameters $u_i,\;u_j$) in  difference form
via the exponential functions,
$\frac{f/h[\varepsilon_i,\mathrm{x}^a_i,\{u_i\}]}
{f/h[\varepsilon_j,\mathrm{x}^a_j,\{u_j\}]}\approx e^{u_i-u_j}$,
and for the argument $u_{ij}=u_i-u_j$ the YB equations have
ordinary form (\ref{rrrsu}). The mentioned property comes from the
fact, that we have obtained the above inhomogeneous solutions as
generalizations to the solutions of the YBE (\ref{rrrsu}).

 But, as we
can see, there is possible to obtain more general inhomogeneous
solutions, where the dependence from the arbitrary functions has
not the discussed "factorization" property. The solutions of the
functions $f_{ij},\; g_{ij},\;h_{ij}$ to the YBE for the
homogeneous cases, i.e. at the values
$\varepsilon_i=\varepsilon_j,\;\mathrm{x}^a_i=\mathrm{x}^a_j$, can
be viewed as primary conditions for the general inhomogeneous
solutions. Further we represent all the constant primary
conditions (constant solutions to YBE), i.e. when also $u_i=u_j$
(spectral parameter dependent ones $u_i \neq u_j$ with YBE
(\ref{rrrsu}) are presented above), and their extensions.

 \paragraph{$\star$}
  The most fruitful case corresponds to  the primary
conditions $f_{ii}=1,\;g_{ii}=h_{ii}=0$. Note that the already
obtained case (\ref{f0}) is one of the such solutions, which has
not included in the three families of the solutions
(\ref{fa1}-\ref{h1}), (\ref{fa2}-\ref{h2}) and
(\ref{f1}-\ref{hf}),  presented in the previous
paragraph. 

%
%
%
%

Hereafter we omit the variables $\mathrm{x}^a$ and $u$ in the
arguments of the functions, as the variables $u_i$ are not
involved immediately in the YBE, and the variables
$\mathrm{x}^a_i$ can be eliminated by the appropriate
normalization of the functions. However, when we obtain a
dependence from an arbitrary function on the parameter
$\varepsilon_i$, then we can involve in the argument the remaining
variables as well.

 We take for the function $f_{ij}$ a general parametrization (\ref{f1})
 \be f_{ij}=\frac{(\mathrm{x}^a_i)^2}{(\mathrm{x}^a_j)^2}
\frac{f[\varepsilon_i]}{f[\varepsilon_j]}.\ee

For presenting the general solutions with the mentioned primary
conditions $f_{ii}=1,\;g_{ii}=h_{ii}=0$ we denote
\bea\nn
 \bar{g}_{ij}=i
 e^{\varepsilon_i}(1+e^{2\varepsilon_j})+\frac{\mathrm{x}^a_j}{\mathrm{x}^a_i}(1+e^{2\varepsilon_i})
 (e^{\varepsilon_i+\varepsilon_j}g_{ij}+h_{ij}),\;\;\;\\
 \bar{h}_{ij}=\frac{f[\varepsilon_j]}{f[\varepsilon_i]}
 \left(e^{\varepsilon_i}\frac{f[\varepsilon_i]}{f[\varepsilon_j]}-e^{\varepsilon_j}
 +i\frac{\mathrm{x}^a_j}{\mathrm{x}^a_i}(e^{\varepsilon_i+\varepsilon_j}h_{ij}+g_{ij})\right).
 \label{hgn}\eea
From the YBE we obtain the following consistency conditions for
the solutions ($g_0$ is a constant)
\bea \label{ggh} \bar{h}_{ij}=0 \;\;\; \textrm{or}\;\;\;
e^{\varepsilon_j}(1+e^{2\varepsilon_i})^2\frac{f[\varepsilon_i]}{f[\varepsilon_j]}+i(1+e^{2\varepsilon_j})\bar{g}_{ij}=
\bar{h}_{ij}\frac {g_0}{(f[\varepsilon_j])^2}.
 \eea

One can unveil the meaning of the above conditions, representing
the $\check{R}$-matrix (\ref{rco}) in terms of the functions $
\bar{h}_{ij},\;\bar{g}_{ij}$. It appears that
$\check{R}_{11}^{00}\approx \bar{h}_{ij}$, and
$\check{R}_{00}^{11}\approx(e^{\varepsilon_j}(1+e^{2\varepsilon_i})^2\frac{f[\varepsilon_i]}{f[\varepsilon_j]}+i(1+e^{2\varepsilon_j})\bar{g}_{ij})
$. Thus the consistency conditions simply imply
$\check{R}_{11}^{00}=0$  or $\check{R}_{00}^{11}=0$ (when
$g_0=0$), or $\check{R}_{00}^{11}\approx
\check{R}_{11}^{00}/({f[\varepsilon_i]}{f[\varepsilon_j]})$.

 At first let us consider the case
$\bar{h}_{ij}=0$. The solutions now have the  forms (the function
$g_{ij}$ can be obtained from the equation (\ref{hgn}))
\bea
f_{ij}=\frac{(\mathrm{x}^a_i)^2}{(\mathrm{x}^a_j)^2}\frac{f[\varepsilon_i]}{f[\varepsilon_j]},\label{hk}
 \;\;\; h_{ij}=i\frac{\mathrm{x}^a_i}{\mathrm{x}^a_j}
\frac{[e^{\varepsilon_j}-\tilde{h}[\varepsilon_j]](1+e^{2\varepsilon_i})
\frac{f[\varepsilon_i]}{f[\varepsilon_j]}-[e^{\varepsilon_i}-\tilde{h}[\varepsilon_i]](1+e^{2\varepsilon_j})}
{(1+e^{2\varepsilon_i})(1+e^{2\varepsilon_j})},
 \eea
where the functions $\tilde{h}[\varepsilon]$ and $f[\varepsilon]$
are interrelated/interdependent. Let $\tilde{h}[\varepsilon]$ is
an arbitrary function, then the  general solutions contain a
constant number $f_0$ and
\bea f[\varepsilon]=\frac{(1+f_0)\tilde{h}[\varepsilon]\pm
\sqrt{(1+f_0^2)\tilde{h}[\varepsilon]^2-2 f_0)}
}{1+e^{2\varepsilon}}. \label{feh}\eea
Of course, 
 one can reverse the
dependence in the relation (\ref{feh}) and write the function
$\tilde{h}[\varepsilon]$ in terms of the arbitrary function
$f[\varepsilon]$, then we shall come to the formula 
\bea h_{ij}=\quad \\\nn
i\frac{\mathrm{x}^a_i}{\mathrm{x}^a_j}\frac{\cosh{[\varepsilon_i]}f[\varepsilon_i]
(1\!\pm \! i \sqrt{e^{-2\varepsilon_j}\!+\!f_0
(\cosh{[\varepsilon_j]}f[\varepsilon_j])^2})\!-\!\cosh{[\varepsilon_j]}f[\varepsilon_j]
(1\!\pm \! i\sqrt{e^{-2\varepsilon_i}
\!+\!f_0(\cosh{[\varepsilon_i]}f[\varepsilon_i])^2}) }
{2f[\varepsilon_j]\cosh{[\varepsilon_i]}\cosh{[\varepsilon_j]}}.
\eea

 When in (\ref{hk}) the function $\tilde{h}[\varepsilon]=0$, then the condition (\ref{feh}) is not
required, the  function $f[\varepsilon]$ is arbitrary, and  we
come to the solution (\ref{f1}, \ref{gf}, \ref{hf}).

Now let us consider the case $\bar{h}_{ij}\neq 0$ in (\ref{ggh}).
When $g_0=0$, then we have the solutions with arbitrary functions
$f[\varepsilon]$ and constant $f_0$:
\bea h_{ij}=
\frac{\mathrm{x}^a_i}{\mathrm{x}^a_j}\frac{i(\cosh{[\varepsilon_i]}f[\varepsilon_i]\!-\!\cosh{[\varepsilon_j]}f[\varepsilon_j])
\!\pm\! \sqrt{f_0
\!+\!e^{2\varepsilon_i}(\cosh{[\varepsilon_i]}f[\varepsilon_i])^2}\!\mp\!
\sqrt{f_0 \!+\!e^{2\varepsilon_j}
(\cosh{[\varepsilon_j]}f[\varepsilon_j])^2}}
{2f[\varepsilon_j]\cosh{[\varepsilon_i]}\cosh{[\varepsilon_j]}}.
\eea

In the case $g_0\neq 0$ the general solutions are of the following
form with arbitrary  $f[\varepsilon]$ and  $g_0$
\bea
\bar{h}_{ij}=\frac{\mathrm{x}^a_i}{\mathrm{x}^a_j}\frac{(1-e^{2(\varepsilon_i+\varepsilon_j)})f[\varepsilon_j]
\Big(\bar{f}[\varepsilon_j]e^{\varepsilon_i}\left(
\bar{f}[\varepsilon_j]\bar{h}[\varepsilon_j]-
\bar{f}[\varepsilon_i]\bar{h}[\varepsilon_i]\right)+ g_0\left(
\bar{f}[\varepsilon_i]\bar{h}[\varepsilon_j]-\bar{f}[\varepsilon_j]
\bar{h}[\varepsilon_i]\right)\Big)}{ g_0(1+\bar{f}[\varepsilon_i]
\bar{f}[\varepsilon_j]\bar{h}[\varepsilon_i]\bar{h}[\varepsilon_j])
+e^{\varepsilon_i}\bar{f}[\varepsilon_j]
\left(\bar{f}[\varepsilon_i] \bar{f}[\varepsilon_j]+g_0^2
\bar{h}[\varepsilon_i]\bar{h}[\varepsilon_j]\right)}\label{rh0}
 \eea
%
%
where $\bar{f}[\varepsilon]=[1+e^{2\varepsilon}]f[\varepsilon]$
and (below $h_0$ is an arbitrary number)
%
%
%
%
\bea \bar{h}[\varepsilon]=
\frac{(\bar{f}[\varepsilon])^2-1}{\bar{f}[\varepsilon]h_0\pm
\sqrt{(\bar{f}[\varepsilon]h_0)^2+((\bar{f}[\varepsilon])^2-g_0^2)
((\bar{f}[\varepsilon])^2-1)}}. \label{fgh}\eea
The functions $h_{ij},\;g_{ij}$ can be obtained then using the
relations (\ref{hgn}) and the second equation in (\ref{ggh}). Let
us remind once again that the arbitrary function $f[\varepsilon]$
can have also an extra argument $u$, and in this case taking the
homogeneous limit $\varepsilon_i=\varepsilon_j,\;
\mathrm{x}^a_i=\mathrm{x}^a_j$ (two identical irreps), but keeping
$u_i\neq u_j$, we shall have two-parametric solution $R(u_i,u_j)$
to YBE.. The solutions (\ref{fa1}-\ref{h1}), (\ref{fa2}-\ref{h2})
obtained previously in this subsection and containing arbitrary
functions $h[\varepsilon_i,x_i^a,\{u_i\}]$ correspond to the
exceptional cases of (\ref{rh0}, \ref{fgh}), with the property
$\bar{f}[\varepsilon]=1,g_0$ ($\bar{f}[\varepsilon]=constant$).

If we impose additional requirements $g_{ij}=0$, $h_{ij}=0$, it
will fix the function $f_{ij}$, as in the case (\ref{f0}). For
completeness, let us present all the particular cases. When
$h_{ij}=0$ and/or $g_{ij}=0$, then under the conditions
$\bar{h}_{ij}\neq 0,\; g_0\neq 0$ (the second relation in
(\ref{ggh})) we shall come to the solution (\ref{f0}). For the
mentioned conditions, there are another particular solutions also:
when $h_{ij}=0$ they are
$f[\varepsilon]=\frac{1}{\cosh{[\varepsilon]}}$ and
$g_{ij}=i\frac{\sinh{[\varepsilon_i-\varepsilon_j]}}{\cosh{[\varepsilon_i]}}$,
when $g_{ij}=0$, the solutions are
$f[\varepsilon]=\frac{e^{-2\varepsilon}}{\cosh{[\varepsilon]}}$
and $h_{ij}=-i
\frac{e^{\varepsilon_j-\varepsilon_i}\sinh{[\varepsilon_i-\varepsilon_j]}}{\cosh{[\varepsilon_i]}}$.
 The  condition $\bar{h}_{ij}=0$ brings to the specific solutions
 $$h_{ij}=0,\;\;\; f[\varepsilon]=\frac{1\pm e^{-\varepsilon}
  \sqrt{f_0 e^{\varepsilon}\cosh{[\varepsilon]}-1}}{\cosh{[\varepsilon]}},\;\;\;
  g_{ij}=i(e^{\varepsilon_i}\frac{f[\varepsilon_i]}{f[\varepsilon_j]}-
  e^{\varepsilon_j})$$ and
$$g_{ij}=0,\;\;\; f[\varepsilon]=\frac{e^{-2\varepsilon}(1\pm\sqrt{1+f_0
e^{\varepsilon}\cosh{[\varepsilon]}})}{\cosh{[\varepsilon]}},\;\;\;
 h_{ij}=i(e^{-\varepsilon_j}\frac{f[\varepsilon_i]}{f[\varepsilon_j]}-
  e^{-\varepsilon_i})$$ The conditions $h_{ij}\neq 0$,
  $g_0=0$ (see the second relation in
(\ref{ggh})) imply
 $$h_{ij}=0,\;\;\; f[\varepsilon]=\frac{e^{-\varepsilon}(1\pm\sqrt{1+f_0
e^{\varepsilon}\cosh{[\varepsilon]}})}{\cosh{[\varepsilon]}^2},\;\;\;
  g_{ij}=i(e^{-\varepsilon_j}\frac{f[\varepsilon_i]\cosh{[\varepsilon_i]}}
  {f[\varepsilon_j]\cosh{[\varepsilon_j]}}-
  e^{-\varepsilon_i}\frac{\cosh{[\varepsilon_j]}}{\cosh{[\varepsilon_i]}})$$ and
$$g_{ij}=0,\;\;\; f[\varepsilon]=\frac{e^{\varepsilon}\pm\sqrt{f_0
e^{\varepsilon}\cosh{[\varepsilon]}-1}}{e^{2\varepsilon}\cosh{[\varepsilon]}^2},\;\;\;
  h_{ij}=i(e^{\varepsilon_i}\frac{f[\varepsilon_i]\cosh{[\varepsilon_i]}}
  {f[\varepsilon_j]\cosh{[\varepsilon_j]}}-
  e^{\varepsilon_j}\frac{\cosh{[\varepsilon_j]}}{\cosh{[\varepsilon_i]}})$$
 As it
was stated the variables $\mathrm{x}^a$ can be eliminated from the
YBE by the appropriate normalization of the functions $f_{ij},\;
g_{ij}$ and $h_{ij}$: $f_{ij}\to
\frac{(\mathrm{x}^a_i)^2}{(\mathrm{x}^a_j)^2}f_{ij}$, $g_{ij}\to
\frac{\mathrm{x}^a_i}{\mathrm{x}^a_j}g_{ij}$ and $h_{ij}\to
\frac{\mathrm{x}^a_i}{\mathrm{x}^a_j} h_{ij}$. For expelling the
variables $\mathrm{x}^a_i$ from the $\check{R}$-matrix
(\ref{rco}), there is need also an additional vector space
renormalization. Hereafter in the formulas we omit the variables
$\mathrm{x}^a$.

\paragraph{$\star\star$}

Next group of the solutions is equipped with the primary
conditions
\bea f_{ii}=1,\;\;\; g_{ii}=-h_{ii}=\pm 1,\\
f_{ii}=1,\;\;\; g_{ii}=h_{ii}=\pm \tanh{[\varepsilon_i]}. \eea

At the first we investigate the case $g_{ii}=h_{ii}=\pm
\tanh{[\varepsilon_i]}$. Here we have two solutions
\bea f_{ij}=\frac{1+e^{2 \varepsilon_j}}{1+e^{2
\varepsilon_i}},\;\;\;
g_{ij}=\frac{\pm(1-e^{\varepsilon_i+\varepsilon_j})+i(e^{
\varepsilon_i}-e^{\varepsilon_j})}{1+e^{2 \varepsilon_i}},\;\;\;
h_{ij}=\frac{\pm(1-e^{\varepsilon_i+\varepsilon_j})-i(e^{
\varepsilon_i}-e^{\varepsilon_j})}{1+e^{2 \varepsilon_i}}. \eea

Then for the case $g_{ii}=-h_{ii}=1$ 
we obtain
\bea f_{ij}=\frac{1+e^{2 \varepsilon_j}}{1+e^{2
\varepsilon_i}},\;\;\; g_{ij}=-\frac{ i-e^{\varepsilon_j}}{i-e^{
\varepsilon_i}},\;\;\; h_{ij}=\frac{ i-e^{\varepsilon_j}}{i-e^{
\varepsilon_i}}. \eea
When  $g_{ii}=-h_{ii}=-1$, then we have
\bea f_{ij}=\frac{1+e^{2 \varepsilon_j}}{1+e^{2
\varepsilon_i}},\;\;\; g_{ij}=\frac{ i+e^{\varepsilon_j}}{i+e^{
\varepsilon_i}},\;\;\; h_{ij}=-\frac{ i+e^{\varepsilon_j}}{i+e^{
\varepsilon_i}}. \eea

\paragraph{$\star\star\star$}  As we can see, the previous group of the solutions exclude the normalization
condition, \\i.e. 
$\check{R}_{ii}(\varepsilon,\varepsilon,\mathrm{x}^a,\mathrm{x}^a,0)\neq\mathbb{I}$.
Another group of such solutions is
\bea f_{ij}=0,\;\;\; g_{ij}=\frac{-i
e^{\varepsilon_j-\varepsilon_i}}{2 \cosh{\varepsilon_i}},\;\;\;
h_{ij}=\frac{-i}{2 \cosh{\varepsilon_i}},\\
f_{ij}=0,\;\;\; g_{ij}=\frac{e^{\varepsilon_j}}{i\pm
e^{\varepsilon_i}},\;\;\; h_{ij}=\frac{-i}{\pm i+e^{\varepsilon_i}},\\
f_{ij}=0,\;\;\;g_{ij}=\frac{\pm 1-i e^{\varepsilon_j}}{1+
e^{2\varepsilon_i}},\;\;\; h_{ij}=\frac{(-i\mp
e^{\varepsilon_j})e^{\varepsilon_i} }{1+e^{2\varepsilon_i}}. \eea

The remaining case can be presented by the following matrix
$\check{R}_{ij}=f_{ij}\breve{P}_{--}+g_{ij}\breve{P}_{+-}+h_{ij}\breve{P}_{-+}$.
 Here the existing solutions to YBE are (we can set $f_{ij}=1$, as there is a
 normalization freedom)
\bea  g_{ij}=\frac{i e^{\varepsilon_-\varepsilon_j}f_{ij}}{2
\cosh{\varepsilon_j}},\;\;\;
h_{ij}=\frac{i f_{ij}}{2 \cosh{\varepsilon_j}},\\
g_{ij}=\frac{-e^{\varepsilon_i}f_{ij}}{i\pm
e^{\varepsilon_j}},\;\;\; h_{ij}=\frac{i f_{ij}}{\pm i+e^{\varepsilon_j}},\\
 g_{ij}=\frac{(i e^{\varepsilon_i}\pm 1)f_{ij}}{1+
e^{2\varepsilon_j}},\;\;\; h_{ij}=\frac{(i\mp e^{\varepsilon_i})
f_{ij}}{1+e^{2\varepsilon_j}}. \eea

So, we exhausted all the possible solutions with the condition
$\mathrm{c}_i=\mathrm{c}_j=0$.

 As we see the solutions with the normalization condition $\check{R}_{ii}=\mathbb{I}$
  contain arbitrary constants and arbitrary functions, forming so
   families of the solutions.
\subsection{$\cosh{\varepsilon}=0$}
 The solutions of the equation $\mathrm{c}_i \cosh{\varepsilon_j}=\pm \mathrm{c}_j \cosh{\varepsilon_i}$
include also the case with $\cosh{\varepsilon_{i,j}}=0$, which we
did not consider above as it corresponds
 to the value  $\mathrm{z_{i,j}}=-1$. But here there is an interesting property of the
 algebra. Two dimensional representation of the algebra now has
 two linearly independent generators, as here $f_{i}=\Big(\mathrm{c}_{i}/\mathrm{x}_{i}\Big) e_{i}$,
 $f_{j}=\Big(\mathrm{c}_{j}/\mathrm{x}_{j}\Big) e_{j}$ and
 $k_{i,j}=diag\{-1,1\}$. But the co-product defined above give good defined four dimensional
 representations for all three generators. The fusion here
 corresponds to the case $V_2 \otimes V_2=V_2\oplus V_2$, where in the summand there are
  two dimensional
 cyclic irreps with the values $e^2=\mathrm{x}=\mathrm{x}_i+\mathrm{x}_j,\; f^2=\mathrm{y}=\mathrm{y}_i+\mathrm{y}_j,\; k^2=1$,
 the quadratic Casimir $c$ has two different
  values on the irreps, differing by the signs,
   $\pm \mathrm{c}_{ij}, \; \mathrm{c}_{ij}=\sqrt{(\mathrm{x}_i+\mathrm{x}_j)(\mathrm{c}_i^2 \mathrm{x}_j+\mathrm{c}_j^2 \mathrm{x}_i)/(\mathrm{x}_i \mathrm{x}_j)}$.
   The
 projection operators $P^{-}_{ij},\; P^{+}_{ij}$ can now be constructed as
 well:
\bea
{\breve{P}}^{\pm}_{ij}=\left(\ba{cccc}\frac{\mathrm{c}_i+\mathrm{c}_j\pm
\mathrm{c}_{ij}}{\mp 2\mathrm{c}_{ij}}&0&0&\frac{\mathrm{c}_i
\mathrm{x}_j-\mathrm{c}_j \mathrm{x}_i}{\pm 2\mathrm{c}_{ij}}
\\0&\frac{\mathrm{c}_i \mathrm{x}_j+\mathrm{c}_j
\mathrm{x}_i}{\mp
2\mathrm{c}_{ij}\mathrm{x}_j}&\frac{\mathrm{c}_j-\mathrm{c}_i\pm
\mathrm{c}_{ij}}{\pm 2\mathrm{c}_{ij}}&0
\\0&\frac{\mathrm{c}_i-\mathrm{c}_j\pm \mathrm{c}_{ij}}{\pm 2\mathrm{c}_{ij}}&\frac{\mathrm{c}_i \mathrm{x}_j+\mathrm{c}_j \mathrm{x}_i}{\mp 2\mathrm{c}_{ij}
\mathrm{x}_i}&0\\\frac{\mathrm{c}_j \mathrm{x}_i-\mathrm{c}_i
\mathrm{x}_j}{\pm 2\mathrm{c}_{ij}\mathrm{x}_i
\mathrm{x}_j}&0&0&\frac{\mathrm{c}_i+\mathrm{c}_j\mp \mathrm{c}_{ij}}{\mp 2\mathrm{c}_{ij}}\ea\right).
\eea
  The simplest matrix
  $\check{R}_{ij}={\breve{P}}^{+}_{ij}+f_{ij}{\breve{P}}^{-}_{ij}$
 satisfying to the YBE is a constant
 {$\scriptsize\left(\ba{cccc}-1&&&\\&&1&\\
 &1&&\\&&&1\ea\right)$} and corresponds to the value $f_{ij}=1$. The next solution has the value $f_{ij}=-1$ and the
 corresponding
 matrix, after multiplication by $-\mathrm{c}_{ij}/(2\sqrt{\mathrm{c}_i \mathrm{c}_j})$ and redefining
 the
 parameters, $\mathrm{c}_{i,j}=e^{2u_{i,j}},\; \mathrm{x}_{i,j}=e^{2w_{i,j}}$, is the
 following (the
 notations $u_{ij}=u_i-u_j$ and $w_{ij}=w_i-w_j$  are used)
 \bea \check{R}(u_{ij},w_{ij})\!=\!\!\!{\small
\left(\!\!\!\ba{cccc}\cosh{[u_{ij}]}&0&0&e^{w_1+w_2}\sinh{[w_{ij}\!-\!u_{ij}]}\\
0&e^{w_{ij}}\cosh{[u_{ij}\!-\!w_{ij}]}&\sinh{[u_{ij}]}&0\\0
 &\sinh{[-\!u_{ij}]}&e^{-w_{ij}}\cosh{[u_{ij}\!-\!w_{ij}]}&0
 \\e^{-\!w_1-\!w_2}\sinh{[u_{ij}\!-\!w_{ij}]}&0&0&\cosh{[u_{ij}]}\ea\!\!\!\right).}
 \label{ruw}\eea

  This is simply the two-parametric solution  \cite{KS}.

  Note, that for the nilpotent or semi-cyclic irreps this
  case equivalent to $c_i=c_j=0$.

\section{Summary}
 \addtocounter{section}{0}\setcounter{equation}{0}

 In this article we have obtained the all spectra of the solutions
 to the $sl_q(2)$-invariant YBE at $q^4=1$ defined on the cyclic irreps. We would like to
 discuss here some peculiarities of the  obtained solutions and the
 corresponding
 integrable models.

We want at first to pay the attention into the following
interesting point regarding to the appearance of the arbitrary
functions in the solutions  of the investigated YBE with
inhomogeneous behavior. The homogeneous choice of the irrep
parameters $\{x_i,z_i,c_i\}=\{x_j,z_j,c_j\}$ then give us
"baxterised" YBE solutions. It is because of the arbitrary
functions. Indeed, as we have noted the arbitrary functions can be
parameterized besides of the irrep characteristics also by some
external parameters: $\{u_i\}, \; \{u_j\}$. Then taking the
homogeneous case in the solutions, we can keep $\{u_i\}\neq
\{u_j\}$, and as result we shall have spectral-parameter dependent
solutions to the homogeneous spectral-parameter dependent YBE. In
this case of the homogeneous limit the arbitrariness of the
functions has not meaning that there is a family of the solutions,
as now the function $f(u_i)$ is just a transformation
(reparametrization) of the spectral parameter $u_i$. We can
separate the obtained solutions as really "baxterised" ones, which
include arbitrary functions, and "just" inhomogeneous solutions,
where all the functions are fixed (the arbitrariness in this case
can be presented only by arising of some constants), and their
homogeneous limits are the solutions to the constant YBE.

As it was observed before the existed $4\times 4$ solutions to
$sl_q(2)$-invariant YBE at $q=\pm i$ all have the "free-fermionic"
property \cite{rbaxter,AuP,RGS,grs,DA}. We can show here that this
is valid for all the solutions on two dimensional cyclic irreps
and moreover: all
 the matrices in the form $\check{R}_{ij}(u)=\sum f_a(u)
 \check{P}_{ij}^a$, with
 the obtained projection matrices (for all three cases discussed in the Section 4),
  independent from the functions $f_a(u)$
 (with arbitrary $f_{a}(u)$), possess
 the following relation on the matrix elements:
\bea \label{free}\check{R}_{00}^{00}(u)\check{R}_{11}^{11}(u)+
\check{R}_{01}^{10}(u)\check{R}_{10}^{01}(u)=
\check{R}_{01}^{01}(u)\check{R}_{10}^{10}(u)+
\check{R}_{00}^{11}(u)\check{R}_{11}^{00}(u). \eea
The chain models corresponding to the obtained solutions all have
the form
 of the $XY$ models in the transverse field. Let us present a
 general expression for the corresponding quantum Hamiltonian operators.
The transfer matrix approach in the theory of the integrable
models implies, that the first logarithmic derivative (at the
normalization point) of the transfer matrix defined on a
one-dimensional chain as $\tau(u)=tr_{j}\prod_i R_{ij}(u)$,
coincides with the Hamiltonian operator of the integrable quantum
spin-chain model, $H=i d \tau(u)/(\tau(u)d u)|_{u=0}$. It means
that the expansion of the $R$-matrix near the point $u=0$, where
$\check{R}(0)=I$, gives interaction terms $h_{i,i+1}$ in the
elementary cell of the nearest-neighbor Hamiltonian operator
$H=\sum_i h_{i,i+1}$. At $u=0$ we have
\bea \check{R}_{00}^{00}(0)=\check{R}_{11}^{11}(0)=
\check{R}_{01}^{01}(0)=\check{R}_{10}^{10}(0)=1 \eea
 and the
remaining elements are vanishing. Expanding near that point the
relation (\ref{free}), we can see, that the derivatives at that
point satisfy to the following relation
\bea {{\check{R}}_{00}^{00}}{}'(0)+{\check{R}_{11}^{11}}{}'(0)=
{\check{R}_{01}^{01}}{ } '(0)+{\check{R}_{10}^{10}}{ }'(0). \eea
The expansion of the $\check{R}(u)$-matrix for the $h_{i,i+1}$ (we
omit the overall coupling constant) gives the following relation,
where we use the Pauli matrices
$\sigma^{+}=(^{0\;1}_{0\;0}),\;\;\;\sigma^{-}=(^{0\;0}_{1\;
0}),\;\;\;\sigma^z=\frac{1}{2}(^{1\;\;\;0}_{0\;{\small -}1})$, 
%
\bea&
h_{i,i+1}=\frac{1}{4}\Big({{\check{R}}_{00}^{00}}{}'(0)+{\check{R}_{11}^{11}}{}'(0)+
{\check{R}_{01}^{01}}{ } '(0)+{\check{R}_{10}^{10}}{
}'(0)\Big)+&\label{hamilt}\\&
\frac{1}{2}\Big({{\check{R}}_{00}^{00}}{}'(0)-{\check{R}_{11}^{11}}{}'(0)+
{\check{R}_{01}^{01}}{ } '(0)-{\check{R}_{10}^{10}}{
}'(0)\Big)\sigma^z_i+\frac{1}{2}\Big({{\check{R}}_{00}^{00}}{}'(0)-{\check{R}_{11}^{11}}{}'(0)-
{\check{R}_{01}^{01}}{ } '(0)+{\check{R}_{10}^{10}}{
}'(0)\Big)\sigma^z_{i+1}+&\nn\\&+
\Big({{\check{R}}_{00}^{00}}{}'(0)+{\check{R}_{11}^{11}}{}'(0)-
{\check{R}_{01}^{01}}{ } '(0)-{\check{R}_{10}^{10}}{ }'(0)\Big)
\sigma^z_i\sigma^z_{i+1}+&\nn\\&\frac{1}{4}\Big({{\check{R}}_{01}^{10}}{}'(0)\sigma^-_i\sigma^+_{i+1}
+{\check{R}_{10}^{01}}{}'(0)\sigma^+_i\sigma^-_{i+1}+
{\check{R}_{00}^{11}}{ }
'(0)\sigma^+_i\sigma^+_{i+1}+{\check{R}_{11}^{00}}{
}'(0)\sigma^-_i\sigma^-_{i+1}\Big).\nn &\eea
We see that the coupling before the interaction term
$\sigma^z_i\sigma^z_{i+1}$ vanishes for the matrices with the
"free-fermionic" property. In terms of the scalar fermions this
summand corresponds to the four fermions' interaction. The
fermionic representation can be performed by the Jordan-Wigner
transformation \cite{grs}, or by a simple method brought in the
work \cite{shs}  for a general $4\times 4$ $\check{R}$-matrix. The
chain models  with the local terms (\ref{hamilt}) in the
Hamiltonian  describe some inhomogeneous $XY$ models in a
transverse field, and the Hamiltonian operators  in the
representation of the fermionic creation and annihilation
operators
  have only  quadratic nearest-neighbored hopping terms.

 As example, the
solution (\ref{ru0}) just describes the $XX$-model in the
transverse magnetic field $\approx\cosh{u_0}$,
\bea H=J\sum_{i}\Big( \sigma_i^+\sigma_{i+1}^-
+\sigma_{i+1}^+\sigma_{i}^-+\cosh{u_0}\sigma_i^z\Big).\eea
Another Hamiltonian operator corresponding to the general solution
(\ref{rzij}), when we normalize the matrix so that $\alpha=0$, and
$f_i/f_j=1$, and the transfer matrix is expanded near the point
$\varepsilon_i=\varepsilon_j=\varepsilon$, is written as
\bea H=J\sum_{i}\Big(
 i(\sigma_i^+\sigma_{i+1}^-
-\sigma_{i+1}^+\sigma_{i}^-)+e^{\varepsilon}(\sigma_{i+1}^z-\sigma_{i}^z)\Big).\eea

Note that this relatively simple free-fermion description arises
 in four-dimensional matrixes case $q=\pm i$. Higher roots of unity,
starting from $9\times9$-dimensional case, corresponding to
$q^3=1$, lead to much more rich variety of solutions and contain
also higher interaction terms in the corresponding one-dimensional
quantum chain Hamiltonian operators. These cases will be
considered elsewhere.

The more interesting results we expect to find are connected with
the cyclic indecomposable representations, as in the case for the
 highest/lowest weight indecomposable representations, considered in \cite{shkh}. There we
 have found solutions, which correspond to the one chain
 Hamiltonian operators with the interactions.

\paragraph{Acknowledgements.} This work is partly supported by
Armenian grant 11-1c028 and by ANSEF grant matph-2908.

\end{document}